\documentclass[aps,prd,showpacs,nofootinbib]{revtex4}
\usepackage{epsfig}
\newcommand{\be}{\begin{equation}}
\newcommand{\ee}{\end{equation}}
\newcommand{\ba}{\begin{eqnarray}}
\newcommand{\ea}{\end{eqnarray}}
\newcommand{\Mn}{M_{\mbox{\tiny N}}}
\newcommand{\pE}{p_{\mbox{\tiny E}}}
\newcommand{\sigmaPiN}{\sigma_{\mbox{\tiny $\pi$N}}}
\newcommand{\trF}{{\rm tr}_{\mbox{\tiny F}}}
\newcommand{\fslash}[1] {{\not\! #1\,}}
\newcommand{\limNR}{\lim\limits_{\mbox{\tiny non-rel}}}
\newcommand{\di}{ {\rm d} }
\newcommand{\la}{\langle}
\newcommand{\ra}{\rangle}
\newcommand{\binomial}[2]{
  \left(\mbox{\small$\!\!{\renewcommand{\arraystretch}{0.8}
  \begin{array}{c} #1 \phantom{|}\!\!  \\ 
                   #2 \phantom{|}\!\! \end{array}}\!\!$}\right)}
\begin{document}
\title{\boldmath 
	The chirally-odd twist-3 distribution function $e^a(x)$ \\
	in the chiral quark-soliton model}
\author{P.~Schweitzer}
\affiliation{Dipartimento di Fisica Nucleare e Teorica, 
  	Universit\`a degli Studi di Pavia, Pavia, Italy}
\date{April 2003}

\begin{abstract}
The chirally-odd twist-3 nucleon distribution $e^a(x)$ is studied in 
the large-$N_c$ limit in the framework of the chiral quark-soliton model 
at a low normalization point of about $0.6\,{\rm GeV}$.
The remarkable result is that in the model $e^a(x)$ 
contains a $\delta$-function-type singularity at $x=0$. 
The regular part of $e^a(x)$ is found to be sizeable at the low scale 
of the model and in qualitative agreement with bag model calculations.
\end{abstract}

\maketitle
\section{Introduction}

In deeply inelastic scattering (DIS) processes the nucleon structure up
to twist-3 is described by six parton distribution functions, the twist-2 
$f_1^a(x)$, $g_1^a(x)$, $h_1^a(x)$ and the twist-3 $e^a(x)$, $g_T^a(x)$, 
$h_L^a(x)$. 
Among these functions the least considered one is probably the twist-3 
chirally odd distribution function $e^a(x)$ \cite{Jaffe:1991kp,Jaffe:1991ra},
which contrasts the fact that it is related to several interesting phenomena.
E.g., a known 
\cite{Balitsky:1996uh,Belitsky:1997zw,Koike:1997bs,Belitsky:1997ay,Kodaira:1998jn}
but rarely emphasized \cite{Efremov:2002qh} fact is that $e^a(x)$ 
contains a $\delta$-function-type singularity at $x=0$, as follows from 
the QCD equations of motion. The existence of a $\delta(x)$-contribution 
also was concluded from perturbative calculations \cite{Burkardt:2001iy}. 
The first Mellin moment of $e^a(x)$ is due to the $\delta(x)$-contribution 
only.
Another interesting phenomenon is connected to the first Mellin moment of 
the flavour-singlet $(e^u+e^d)(x)$ which is proportional to the pion-nucleon
sigma-term $\sigmaPiN$.
The latter gives rise to the so-called ``sigma-term puzzle''. 
The large value extracted from pion-nucleon scattering data,
$\sigmaPiN\approx(50-70)\,{\rm MeV}$ \cite{Koch:pu,Pavan:2001wz}, implies 
that about $20\%$ of the nucleon mass $\Mn$ is due to the strange quark 
-- an unexpectedly large number from the point of view of the OZI-rule.

The reason why $e^a(x)$ has received only little attention so far is related 
to its chiral odd nature, which means that $e^a(x)$ can enter an observable 
only in connection with another chirally odd distribution or fragmentation 
function, and therefore is difficult to access in experiments. 
Only recently it was shown that $e^a(x)$ can be accessed by means of 
the ``Collins effect'' \cite{collins}, i.e.\  the left-right asymmetry in 
the fragmentation of a transversely polarized quark into a pion.
This effect is described by the chirally and T-odd fragmentation function 
$H_1^{\perp a}(z)$, which is ``twist-2'' in the sense that its contribution 
to an observable is not power suppressed \cite{collins,muldt}.
First experimental indications to $H_1^\perp$ were reported in
\cite{Efremov:1998vd}.
Assuming factorization it was shown that the Collins effect gives 
rise to a specific azimuthal 
(with respect to the axis defined by the exchanged hard virtual photon) 
distribution of pions produced in DIS of longitudinally polarized electrons 
off an unpolarized proton target. The observable single (beam) spin asymmetry
is proportional to $\sum_a e_a^2 \,e^a(x)\,H_1^{\perp a}(z)$ \cite{Mulders:1996dh}
($e_a=\pm\frac23,\pm\frac13$ are the quark electric charges).
The process was studied in the HERMES experiment and the effect found 
consistent with zero within error bars \cite{hermes}.\footnote{
	\label{footnote-1}
	The prominent result of the HERMES experiment \cite{hermes} is the
	observation of sizeable azimuthal asymmetries in pion production from 
	DIS of {\sl unpolarized electrons} off {\sl longitudinally polarized 
	protons}, which contain information on $H_1^{\perp a}(z)$ and the 
	chirally odd distribution functions $h_1^a(x)$ and $h_L^a(x)$ 
	\cite{Mulders:1996dh}.}
However, in the CLAS  experiment, in a different kinematics, 
a sizeable asymmetry was observed \cite{Avakian-talk,Avakian:2003pk}.
If the interpretation applies, that the CLAS data 
\cite{Avakian-talk,Avakian:2003pk} are due to the Collins effect, 
then $e^a(x)$ is definitely not small.
Using estimates of $H_1^\perp(z)$ from HERMES data \cite{Efremov:2001cz}
it was shown in \cite{Efremov:2002sd}, that $e^a(x)$ could be about
half the magnitude of the unpolarized twist-2 distribution $f_1^a(x)$ 
at $Q^2 \sim 1.5 \,{\rm GeV}^2$ in the region $0.15\le x\le 0.4$ covered 
in the CLAS experiment.

The indication that $e^a(x)$ could be large in the valence-$x$ region 
is not surprising, if one considers results from the bag model 
\cite{Jaffe:1991ra,Signal:1997ct}, the only model where $e^a(x)$ has 
been studied so far.  In this note $e^a(x)$ will be studied in the 
chiral quark-soliton model ($\chi$QSM).
A subtle question is whether models with no gluon degrees of freedom
(bag model, $\chi$QSM) can describe twist-3 distribution functions.
The answer given in \cite{Jaffe:1991kp,Jaffe:1991ra} is yes, because 
$e^a(x)$, $g_T^a(x)$ and $h_L^a(x)$ are special cases of more general 
quark-gluon-quark correlation functions; special inasmuch they do not 
contain {\sl explicit} gluon fields. 
However, implicitly gluons do contribute and it is important and instructive 
to carefully interpret the results.  E.g., in the bag model $e^a(x)$ is 
due to the bag boundary \cite{Jaffe:1991ra}.
This can be understood considering that the bag boundary (in a 
most intuitive way) models confinement, and thus ``mimics'' gluons.

The $\chi$QSM was derived from the instanton model of the QCD vacuum.
An important small parameter in this derivation is the ``instanton packing
fraction'' which characterizes the diluteness of the instanton medium
\cite{Diakonov:1985eg,Diakonov:1983hh}.
Gluon degrees of freedom appear only at next-to-leading order of this
parameter \cite{Diakonov:1995qy}.
In leading order of the instanton packing fraction the $\chi$QSM quark
degrees of freedom can be identified with the QCD quark degrees of freedom.
This allows to consistently describe twist-2 quark and antiquark distribution 
functions $f_1^a(x)$, $g_1^a(x)$ and $h_1^a(x)$ at a low scale around 
$600\,{\rm MeV}$ \cite{Diakonov:1996sr}. The numerical results 
\cite{Diakonov:1996sr,Diakonov:1997vc,Weiss:1997rt,Pobylitsa:1996rs,Goeke:2000wv} 
agree within $(10-30)\%$ with parameterizations for the ``known'' 
distribution functions performed at low scales \cite{GRV+GRSV}.

The twist-3 distribution functions $g_T^a(x)$ and $h_L^a(x)$ were studied 
in the instanton vacuum model in Refs.~\cite{Balla:1997hf,Dressler:2000hc}.
The remarkable conclusion was that the pure twist-3 interaction dependent 
parts $\widetilde{g}_T^a(x)$ and $\widetilde{h}_L^a(x)$ in the
Wandzura-Wilczek(-like) decompositions of $g_T^a(x)$ and $h_L^a(x)$ 
are strongly suppressed by powers of the instanton packing fraction.
As $g_T^a(x)$ and $h_L^a(x)$ do not contain {\sl explicit} gluon fields it is
possible to evaluate $g_T^a(x)$ and $h_L^a(x)$ directly in the $\chi$QSM.
This was done in Refs.~\cite{Wakamatsu:2000ex,Wakamatsu:2001fd},
and it was observed that the pure twist-3 parts $\widetilde{g}_T^a(x)$
and $\widetilde{h}_L^a(x)$ are indeed small. 
Thus in these cases the $\chi$QSM respects the results found directly 
in the instanton vacuum model -- i.e.\  the theory from which it was
derived. This experience encourages to also tackle the study of $e^a(x)$ in
the $\chi$QSM. However, one has to keep in mind that the results and their 
interpretation presented here should also be reexamined in the instanton 
vacuum model. This is out of the scope of this note and left for future 
studies.

The $\chi$QSM describes the nucleon as a chiral soliton of the pion field in
the limit of a large number of colours $N_c$. This note focuses on the 
flavour singlet distributions $(e^u+e^d)(x)$ and $(e^{\bar u}+e^{\bar d})(x)$
-- the leading flavour combinations in the large-$N_c$ limit.
The consistency of the approach is checked, by demonstrating that 
the model expressions for $(e^u+e^d)(x)$ satisfy QCD sum rules. 
It is shown that the model expressions are (quadratically and 
logarithmically) UV-divergent, and a consistent regularization is defined.
Remarkably, it is found that the $\chi$QSM-expression for 
$(e^u+e^d)(x)$ contains a $\delta(x)$-contribution.
The UV-behaviour and the $\delta(x)$-contribution make an exact numerical 
evaluation of $(e^u+e^d)(x)$ and $(e^{\bar u}+e^{\bar d})(x)$ particularly 
involved. Therefore $(e^u+e^d)(x)$ is evaluated using an approximation, 
the ``interpolation formula'' of Ref.~\cite{Diakonov:1996sr}.

This note is organized as follows.  In Sec.~\ref{Sec-e-in-theory} the twist-3 
distribution function $e^a(x)$ and some of its properties are discussed.
Sec.~\ref{Sec-model} contains a brief introduction into the $\chi$QSM.
In Sec.~\ref{Sec-e-in-model} the flavour-singlet distribution $(e^u+e^d)(x)$
is discussed and evaluated in the model using the interpolation formula.
In Sec.~\ref{App-non-rel-limit} $e^a(x)$ is discussed in the 
non-relativistic limit.
Sec.~\ref{conclusions} contains a summary and conclusions.

\section{\boldmath The distribution function $e^a(x)$}
\label{Sec-e-in-theory}

The chirally odd twist-3 distribution functions $e^q(x)$ for quarks 
of flavour $q$ and $e^{\bar q}(x)$ for antiquarks of flavour $\bar q$ 
are defined as \cite{Jaffe:1991kp,Jaffe:1991ra}
\be\label{def-e}
	e^q(x) = \frac{1}{2\Mn} \int\!\frac{\di\lambda}{2\pi}\,e^{i\lambda x}\,
	\la N|\,\bar{\psi}_q(0)\,[0,\lambda n]\,\psi_q(\lambda n)\,|N\ra 
	\;,\;\;\; e^{\bar q}(x)=e^q(-x) \;,\ee
where $[0,\lambda n]$ denotes the gauge-link. 
The scale dependence is not indicated for brevity. 
The light-like vectors $n^\alpha$ in Eq.~(\ref{def-e}) and $p^\alpha$ are 
defined such that $n^\alpha p_\alpha=1$ and the nucleon momentum is given 
by $P^\alpha_{\mbox{\tiny N}}=n^\alpha+p^\alpha \Mn^2/2$.
The matrix element in Eq.~(\ref{def-e}) is averaged over nucleon spin,
i.e. $\la N|\dots|N\ra \equiv \frac12\sum_{S_3}\la N,S_3|\dots|N,S_3\ra$.

The renormalization scale evolution of $e^a(x)$ was studied 
in Refs.~\cite{Balitsky:1996uh,Belitsky:1997zw,Koike:1997bs},
see also Refs.~\cite{Belitsky:1997ay,Kodaira:1998jn} for reviews.
It evolves according to an evolution pattern typical for twist-3 quantities.
It is not sufficient to know the $n^{\rm th}$ Mellin-moment 
${\cal M}_n[e^a](Q_0^2) = \int\di x\,x^{n-1}e^a(x,Q_0^2)$ at an initial scale 
$Q_0^2$, in order to compute ${\cal M}_n[e^a](Q^2)$ for $Q^2 > Q_0^2$.
Instead, the knowledge of all moments ${\cal M}_k[e^a](Q^2_0)$ 
with $k\le n$ is required.
In the limit of a large number of colours $N_c$ the evolution of $e^a(x)$ 
simplifies to a DGLAP-type evolution -- as it does for the other two proton 
twist-3 distributions $h_L^a(x)$ and (the flavour non-singlet) $g_T^a(x)$.

The QCD-equations of motion allow to decompose $e^q(x)$ in a gauge-invariant 
way as \cite{Balitsky:1996uh,Belitsky:1997zw,Koike:1997bs}, see also 
\cite{Belitsky:1997ay,Kodaira:1998jn} and \cite{Efremov:2002qh},
\be\label{e-decomposition}
	e^q(x) = \delta(x)\,\frac{1}{2\Mn}\,\la N|\bar\psi_q(0)\psi_q(0)|N\ra 
	       + e^q_{tw3}(x)
	       + \frac{m_q}{\Mn}\;\frac{f_1^q(x)}{x}\;\delta_{n>1}\;\;.\ee
The $\delta(x)$-contribution has no partonic interpretation. 
Some authors cancel it out by multiplying $e^q(x)$ by $x$, while others
prefer to consider from the beginning an alternative definition of $e^q(x)$ 
with an explicit factor of $x$ on the RHS of Eq.~(\ref{def-e}).
The existence of a $\delta(x)$-contribution also was concluded 
in Ref.~\cite{Burkardt:2001iy}, where $e^q(x)$ was constructed 
explicitly for a one-loop dressed massive quark.
The contribution $e^q_{tw3}(x)$ in 
Eq.~(\ref{e-decomposition}) is a quark-gluon-quark correlation function, i.e. 
the actual ``pure'' twist-3 (``interaction dependent'') contribution to 
$e^a(x)$, and has a partonic interpretation as an interference between 
scattering from a coherent quark-gluon pair and from a single quark 
\cite{Jaffe:1991kp,Jaffe:1991ra}. Its first two moments vanish.
The third (``mass''-) term in Eq.~(\ref{e-decomposition}) vanishes in the 
chiral limit. The ``Kronecker symbol'' $\delta_{n>1}$ accompanying it has 
the following meaning. The first moment of the mass-term  vanishes.
For $x\neq 0$ the expression in Eq.~(\ref{e-decomposition}) is, however,
correct and can be used literally to take higher moments $n>1$. 

The first and the second moment of $e^q(x)$ satisfy the sum rules
\cite{Jaffe:1991kp,Jaffe:1991ra} 
\ba\label{e-1moment}
	\int_{-1}^1\di x\;e^q(x) &=& 
	\frac{1}{2\Mn}\la N|\,\bar{\psi}_q(0)\psi_q(0)\,|N\ra\;,\\
\label{e-2moment}
	\int_{-1}^1\di x\;x\,e^q(x) &=& \frac{m_q}{\Mn}\; N_q \;. \ea
In Eq.~(\ref{e-2moment}) $N_q$ denotes the number of the respective 
valence quarks (for proton $N_u=2$ and $N_d=1$).
The sum rules (\ref{e-1moment},~\ref{e-2moment}) follow immediately 
from the decomposition in Eq.~(\ref{e-decomposition}) and the 
above mentioned properties of $e^q_{tw3}(x)$ and the mass-term.
In particular, the sum rule (\ref{e-1moment}) is saturated by 
the $\delta(x)$-contribution in Eq.~(\ref{e-decomposition}).

The flavour singlet distribution function $(e^u+e^d)(x)$ is related 
to the scalar isoscalar nucleon form-factor $\sigma(t)$ defined as
\ba\label{sigma-t}
	\sigma(t)\;\bar{u}_N({\bf P'}) u_N({\bf P}) = 
	m\;\la N({\bf P'})|\,
	\biggl(\bar{\psi}_u(0)\psi_u(0)+\bar{\psi}_d(0)\psi_d(0)\biggr)\,
	|N({\bf P})\ra \;\;,\;\;\; t = (P'-P)^2 \;, \ea
where $u_N$ denotes the nucleon spinor (normalized as $\bar{u}_Nu_N=2\Mn$) 
and $m=\frac12\,(m_u+m_d)$. In (\ref{sigma-t}) a term proportional 
to $(m_u-m_d)(\bar{\psi}_u\psi_u-\bar{\psi}_d\psi_d)$ is neglected.
At the point $t=0$ the form-factor $\sigma(t)$ is referred to as
the pion-nucleon sigma-term $\sigmaPiN$ and related to $(e^u+e^d)(x)$ 
by means of the sum rule (\ref{e-1moment}) as
\be\label{e-1moment-a}
	\sigmaPiN\equiv\sigma(0)=m\int_{-1}^1\di x\;(e^u+e^d)(x) \;\;.\ee
The form factor $\sigma(t)$ describes the elastic scattering off the 
nucleon due to the exchange of a spin-zero particle and is not measured yet.
However, low-energy theorems allow to deduce its value at the Cheng-Dashen 
point $t=2 m_\pi^2$ from pion-nucleon scattering data. One finds
\be\label{sigma-at-Cheng-Dashen-point}
	\sigma(2m_\pi^2) = \cases{
	(64\pm 8)\,{\rm MeV} & Ref.~\cite{Koch:pu}\cr
	(79\pm 7)\,{\rm MeV} & Ref.~\cite{Pavan:2001wz}.}\ee
The difference $\sigma(2m_\pi^2)-\sigma(0)$ was obtained from a 
dispersion relation analysis \cite{Gasser:1990ce} and chiral perturbation 
theory calculations \cite{Becher:1999he} with the consistent result of
$14\;{\rm MeV}$. This yields 
\be\label{sigmaPiN}
	\sigmaPiN \approx (50-70)\;{\rm MeV} \;. \ee
With $m \approx (5-8)\,{\rm MeV}$ one obtains a large number 
for the first moment of $(e^u+e^d)(x)$
\be\label{e-1moment-b}
	\int_{-1}^1\di x\;(e^u+e^d)(x) \approx 10 \;.
\ee
One should keep in mind, however, that the large number in 
Eq.~(\ref{e-1moment-b}) is not due to a large ``valence'' structure 
in $(e^u+e^d)(x)$, but solely due to the $\delta(x)$-contribution.

\section{The chiral quark soliton model \boldmath ($\chi$QSM)}
\label{Sec-model}

The $\chi$QSM is based on the effective chiral relativistic 
quantum field theory given by the partition function 
\cite{Diakonov:1985eg,Diakonov:1987ty,Diakonov:tw}
\ba\label{eff-theory}
&&	Z_{\rm eff} = \int\!\!{\cal D}\psi\,{\cal D}\bar{\psi}\,{\cal D}U\;
	\exp\Biggl(i\int\!\!\di^4x\;\bar{\psi}\,
	(i\fslash{\partial}-M\,U^{\gamma_5}-m)\psi\Biggr) \;\;,\\
\label{def-U}
&&	U=\exp(i\tau^a\pi^a) \;\;,\;\;\;
	U^{\gamma_5} = \frac12(U+U^\dag)+\frac12(U-U^\dag)\gamma_5\;.\ea
In Eq.~(\ref{eff-theory}) $M$ is the (generally momentum dependent) dynamical
quark mass, which is due to spontaneous breakdown of chiral symmetry.
$U$ denotes the $SU(2)$ chiral pion field. 
The current quark mass $m$ in Eq.~(\ref{eff-theory}) explicitly breaks
the chiral symmetry and can be set to zero in many applications.
For certain quantities, however, it is convenient or even necessary 
to consider finite $m$.
The effective theory (\ref{eff-theory}) contains the Wess-Zumino term  
and the four-derivative Gasser-Leutwyler terms with correct coefficients.
It has been derived from the instanton model of the QCD vacuum 
\cite{Diakonov:1983hh,Diakonov:1987ty}, and is valid at low energies 
below a scale set by the inverse of the average instanton size
\be\label{scale}
	\rho_{\rm av}^{-1} \approx 600\,{\rm MeV} \;. \ee
In practical calculations it is convenient to take the momentum dependent 
quark mass constant, i.e. $M(p)\to M(0) = 350\,{\rm MeV}$. In this case 
$\rho_{\rm av}^{-1}$ is to be understood as the cutoff, at which quark 
momenta have to be cut off within some appropriate regularization scheme.
It is important to remark that $(M\rho_{\rm av})^2$ is proportional to the 
parametrically small instanton packing fraction
\be\label{packing-fraction}
	(M\rho_{\rm av})^2 \propto 
	\biggl(\frac{\rho_{\rm av}}{R_{\rm av}}\biggr)^{\!4} \ll 1 \;,\ee
with $R_{\rm av}$ denoting the average distance between instantons.
The smallness of this quantity has been used in the derivation of the
effective theory (\ref{eff-theory}) from the instanton vacuum model
\cite{Diakonov:1983hh,Diakonov:1987ty}.

The $\chi$QSM is an application of the effective theory (\ref{eff-theory}) 
to the description of baryons \cite{Diakonov:1987ty}.
The large-$N_c$ limit allows to solve the path integral over pion field 
configurations in Eq.~(\ref{eff-theory}) in the saddle-point approximation.
In the leading order of the large-$N_c$ limit the pion field is static, and 
one can determine the spectrum of the one-particle Hamiltonian of the 
effective theory (\ref{eff-theory})
\be\label{Hamiltonian}
	\hat{H}|n\ra=E_n |n\ra \;,\;\;
	\hat{H}=-i\gamma^0\gamma^k\partial_k+\gamma^0MU^{\gamma_5}+\gamma^0m 
	\;. \ee
The spectrum consists of an upper and a lower Dirac continuum, distorted by 
the pion field as compared to continua of the free Dirac-Hamiltonian
\be\label{free-Hamiltonian} 
	\hat{H}_0|n_0\ra = E_{n_0}|n_0\ra \;\;,\;\;\;\; 
	\hat{H}_0 = -i\gamma^0\gamma^k\partial_k+\gamma^0 M+\gamma^0m\;, \ee
and of a discrete bound state level of energy $E_{\rm lev}$, 
if the pion field is strong enough.
By occupying the discrete level and the states of the lower continuum each 
by $N_c$ quarks in an anti-symmetric colour state, one obtains a state 
with unity baryon number. 
The soliton energy $E_{\rm sol}$ is a functional of the pion field
\be\label{soliton-energy}
	E_{\rm sol}[U] = N_c \biggl(E_{\rm lev}+
	\sum\limits_{E_n<0}(E_n-E_{n_0})\biggr)\biggr|_{\rm reg} \;. \ee 
$E_{\rm sol}[U]$ is logarithmically divergent and has to be regularized 
appropriately, as indicated in (\ref{soliton-energy}).
Minimization of $E_{\rm sol}[U]$ determines the self-consistent soliton 
field $U_c$.  This procedure is performed for symmetry reasons in the 
so-called hedgehog ansatz 
\be\label{hedgehog}
	\pi^a({\bf x})= e_r^a\;P(r) \;,\;\; 
	U({\bf x})=\cos P(r)+i \tau^a e_r^a \sin P(r)\;,\ee
with $r=|{\bf x}|$ and $e_r^a = x^a/r$, in which the variational problem 
reduces to the determination of the self-consistent soliton profile $P_c(r)$.
The nucleon mass $\Mn$ is given by $E_{\rm sol}[U_c]$.
The momentum and the spin and isospin quantum numbers of the baryon 
are described by considering zero modes of the soliton.
Corrections in the $1/N_c$-expansion can be included by considering 
time dependent pion field fluctuations around the solitonic solution.
A good and for many purposes sufficient approximation to the self-consistent 
profile $P_c(r)$ is given by the analytical ``arctan-profile'' 
\be\label{arctan-profile}
	P(r,m_\pi) = -2\,{\rm arctan}\biggl(\frac{R^2_{sol}}{r^2}(1+m_\pi r)
	e^{-m_\pi r} \biggr)\;,\;\; R_{sol} = M^{-1} \;, \ee
where $R_{sol}$ is the soliton size. In the chiral limit $m\to 0$ in 
(\ref{eff-theory}) the self-consistent profile $P_c(r)\propto 1/r^2$ for 
$r\to\infty$, and then $m_\pi=0$ in (\ref{arctan-profile}). 
If $m\neq 0$ in (\ref{eff-theory}) the self-consistent profile exhibits
a Yukawa-tail $P_c(r)\propto e^{-m_\pi r}/r$ for $r\to\infty$ and $m_\pi$
is the physical pion mass connected to the current quark mass $m$ in 
(\ref{eff-theory}) by the Gell-Man--Oakes--Renner relation 
(see below Eq.~(\ref{Gell-Mann--Oakes--Renner-relation})). 
The analytical profile (\ref{arctan-profile}) simulates this 
and the Gell-Mann--Oakes--Renner relation holds approximately.

The $\chi$QSM allows to evaluate in a parameter-free way nucleon 
matrix elements of QCD quark bilinear operators as (schematically)
\ba	
	\la N|\bar{\psi}(z_1)\Gamma\psi(z_2)|N\ra
	 =  c_{\mbox{\tiny$\Gamma$}} 2\Mn N_c \sum\limits_{n\,\rm occ}
	\int\!\!\di^3{\bf X}\: \bar{\Phi}_n({\bf z}_1-{\bf X})\Gamma
	\Phi_n({\bf z}_2-{\bf X})\, e^{iE_n(z^0_1-z^0_2)} + \dots\;\, 
	\label{matrix-elements-occ} && \\
	 = -c_{\mbox{\tiny$\Gamma$}} 2\Mn N_c \sum\limits_{n\,\rm non}
	\int\!\!\di^3{\bf X}\: \bar{\Phi}_n({\bf z}_1-{\bf X})\Gamma
	\Phi_n({\bf z}_2-{\bf X})\, e^{iE_n(z^0_1-z^0_2)} + \dots\;. 
	\label{matrix-elements-non} && \ea
In Eqs.~(\ref{matrix-elements-occ},~\ref{matrix-elements-non}) $\Gamma$ 
is some Dirac- and flavour-matrix, $c_{\mbox{\tiny$\Gamma$}}$ a constant 
depending on $\Gamma$ and the spin and flavour quantum numbers of the nucleon 
state $|N\ra=|S_3,T_3\ra$, and $\Phi_n({\bf x}) = \la{\bf x}|n\ra$ are the 
coordinate space wave-functions of the single quark states $|n\ra$ 
in (\ref{Hamiltonian}).
The sum in Eq.~(\ref{matrix-elements-occ}) goes over occupied levels $n$
(i.e. $n$ with $E_n\le E_{\rm lev}$), and vacuum subtraction is 
implied for $E_n < E_{\rm lev}$ analogue to Eq.~(\ref{soliton-energy}).
The sum in Eq.~(\ref{matrix-elements-non}) goes over non-occupied levels $n$
(i.e. $n$ with $E_n > E_{\rm lev}$), and vacuum subtraction is 
implied for all $E_n > E_{\rm lev}$.\footnote{
	\label{footnote-equivalence}
	The possibility of computing model expressions in the two ways 
	-- (\ref{matrix-elements-occ}) or (\ref{matrix-elements-non}) -- 
	has a deep connection to the analyticity and locality properties of 
	the model \cite{Diakonov:1996sr}. In practice it provides a powerful 
	check of numerical results.} 
The dots in Eqs.~(\ref{matrix-elements-occ},~\ref{matrix-elements-non}) 
denote terms subleading in the $1/N_c$-expansion, which will not be needed 
in this work. Depending on the Dirac- and flavour-structure the expressions 
(\ref{matrix-elements-occ},~\ref{matrix-elements-non}) can possibly be 
UV-divergent and need to be regularized.  
If in QCD the quantity on the LHS of Eq.~(\ref{matrix-elements-occ}) is 
normalization scale dependent, the model results refer to a scale 
roughly set by $\rho_{\rm av}^{-1}$ in Eq.~(\ref{scale}).

In the way sketched in (\ref{matrix-elements-occ},~\ref{matrix-elements-non})
static nucleon properties (form-factors, axial properties, etc., see 
\cite{Christov:1995vm} for a review), twist-2 
\cite{Diakonov:1996sr,Diakonov:1997vc,Weiss:1997rt,Pobylitsa:1996rs,Goeke:2000wv} 
and twist-3 \cite{Wakamatsu:2000ex,Wakamatsu:2001fd} quark and antiquark 
distribution functions, and off-forward distribution functions 
\cite{Petrov:1998kf} have been studied in the $\chi$QSM.
As far as those quantities are known, the model results agree 
within $(10-30)\%$ with experiment or phenomenology.
It is important to note the theoretical consistency of the approach, 
in particular the quark and antiquark distribution functions in the model 
satisfy all general QCD requirements (sum rules, positivity, inequalities, 
etc.). 

\section{\boldmath $e^a(x)$ in the $\chi$QSM}
\label{Sec-e-in-model}

\subsection{Expressions and consistency}
\label{Subsec-expressions-consistency}

\paragraph{Model expressions.}
The model expressions for the flavour combinations $(e^u\pm e^d)(x)$ 
``follow'' from the expressions for the unpolarized twist-2 distributions 
$(f_1^u\pm f_1^d)(x)$ derived in Ref.~\cite{Diakonov:1996sr} by ``replacing'' 
the relevant Dirac-structure $n^\alpha\gamma_\alpha$ in the definition of
$f_1^a(x)$ by $1/\Mn$. This can be checked by an explicit calculation
which closely follows the derivation given in \cite{Diakonov:1996sr}
and can therefore be skipped here.
This ``analogy'' between $e^a(x)$ and $f_1^a(x)$ is due to the fact that 
both are ``spin average'' distributions, and the relevant Dirac- and
flavour-structures exhibit the same properties under the hedgehog symmetry 
transformations. 
As a consequence, the flavour combinations $(e^u\pm e^d)(x)$ have the same 
large-$N_c$ behaviour as $(f_1^u\pm f_1^d)(x)$ \cite{Diakonov:1996sr}, namely
\ba\label{large-Nc}
	(e^u+e^d)(x) &=& N_c^2\:d(N_c x) \nonumber\\
	(e^u-e^d)(x) &=& N_c\,\:d(N_c x) \;,\ea  
where the functions $d(y)$ are stable in the limit $N_c\to\infty$ for fixed 
arguments $y=N_cx$, and different for the different flavour combinations.
Though derived in the $\chi$QSM, the relations in (\ref{large-Nc}) are of 
general character, considering that the $\chi$QSM is a particular realization 
of the large-$N_c$ picture of the nucleon \cite{Witten:1979kh}.
The relations (\ref{large-Nc})  are already the end of the story of 
``analogies'' between $e^a(x)$ and $f_1^a(x)$ in a relativistic model.
In the non-relativistic limit, however, $e^a(x)$ and $f_1^a(x)$
become equal, see Sec.~\ref{App-non-rel-limit} below.

In this work only the leading order in the large-$N_c$ limit 
will be considered. At this order the model expressions read
\ba\label{mod-e-occ}
	(e^u+e^d)(x) 
	&=& N_c \Mn \sum\limits_{n\,\rm occ} 
	    \la n|\gamma^0\delta(x\Mn-\hat{p}^3-E_n) |n\ra\\
\label{mod-e-non}
	&=& -\, N_c \Mn \sum\limits_{n\,\rm non} 
	    \la n|\gamma^0\delta(x\Mn-\hat{p}^3-E_n) |n\ra \ea
and $(e^u-e^d)(x) = 0$ as anticipated in (\ref{large-Nc}).

\paragraph{Sum rule for the first moment.}
The first moment of the model expression in Eq.~(\ref{mod-e-occ}) reads
\be\label{mod-e-1moment-check1}
	\int\limits_{-1}^1\!\di x\;(e^u+e^d)(x)
	= N_c \sum\limits_{n\,\rm occ} \la n|\gamma^0 |n\ra 
	\equiv \frac{\sigmaPiN}{m} \;. \ee
When integrating over $x$ in (\ref{mod-e-1moment-check1}) one can substitute
$x\to y = x\Mn$ and extend the $y$-integration range $[-\Mn,\Mn]$ to the whole
$y$-axis in the large $N_c$-limit.
The final step in (\ref{mod-e-1moment-check1}) follows by recognizing in 
the intermediate step in (\ref{mod-e-1moment-check1}) the model expression 
for the scalar isoscalar form-factor $\sigma(t)$
\be\label{sigma-t-mod}
	\sigma(t) 
	= m\;N_c\int\!\!\di^3{\bf x}\;J_0(\sqrt{-t}\,|{\bf x}|) 
	\sum\limits_{n\,\rm occ}\Phi^\ast_n({\bf x})\gamma^0 \Phi_n({\bf x})
	\;,\ee
at $t=0$. (The Bessel-function $J_0(z)=\frac{\sin z}{z}\to 1$ for $z\to 0$.) 
The pion-nucleon sigma-term $\sigmaPiN=\sigma(0)$ was studied in the 
$\chi$QSM in \cite{Diakonov:1988mg} and the form-factor $\sigma(t)$ 
in \cite{Kim:1995hu}.

The model expression for $\sigmaPiN$ can also be derived in an 
alternative way using the Feynman-Hellmann theorem (this method 
was used in \cite{Diakonov:1988mg})
\be\label{sigma-Mn-relation}
        \sigmaPiN = m \;\frac{\partial\Mn(m)}{\partial m\;} \;. \ee
Rewriting the expression for the nucleon mass $\Mn\equiv\Mn(m)$ in 
Eq.~(\ref{soliton-energy}) as 
\be\label{mod-e-1moment-check2}
	\Mn(m) = N_c\sum_{n\,\rm occ} E_n 
	=  N_c\sum_{n\,\rm occ} \la n|\hat{H}|n\ra 
	\equiv \Mn(0) + m\;N_c \sum_{n\,\rm occ} \la n|\gamma^0|n\ra \;,\ee
where vacuum subtraction is implied, and inserting $\Mn(m)$ 
(\ref{mod-e-1moment-check2}) into (\ref{sigma-Mn-relation}) 
one recovers the model expressions for $\sigmaPiN$ in 
Eqs.~(\ref{mod-e-1moment-check1},~\ref{sigma-t-mod}).
%
%
This proof is formally correct but one should be careful about
regularization. A comment on that will be made at the end of 
Section~\ref{Subsec-discuss-results}.

\paragraph{Sum rule for the second moment.}
The second moment of $(e^u+e^d)(x)$ in Eq.~(\ref{mod-e-occ}) is
\be\label{mod-e-2moment-check0}
	\int\limits_{-1}^1\!\di x\;x\,(e^u+e^d)(x) 
	= \frac{N_c}{\Mn} \sum\limits_{n\,\rm occ} 
	  \la n|\gamma^0(\hat{p}^3+E_n)|n\ra
	= \frac{N_c}{\Mn} \sum\limits_{n\,\rm occ} 
	  E_n\,\la n|\gamma^0|n\ra \;, \ee
where $\la n|\gamma^0 \hat{p}^3|n\ra$ drops out due to hedgehog symmetry.
In QCD the sum rule (\ref{e-2moment}) follows from using equations of motion. 
In the model the analogon is to use $E_n|n\ra = \hat{H}|n\ra$. One obtains
\ba\label{mod-e-2moment-check1}
&&	\int\limits_{-1}^1\!\di x\;x\,(e^u+e^d)(x) 
	= \frac{N_c}{\Mn}(m + \beta M) \;, \\
\label{mod-e-2moment-check1a}
&&	\beta \equiv \sum\limits_{n\,\rm occ}
	\la n|\,\frac{U+U^\dag}{2}\,|n\ra\;, \ea
where the relation 
$E_n\,\la n|\gamma^0|n\ra = \frac12\,\la n|\{\hat{H},\gamma^0\}|n\ra$
$= m + M \la n|\,\frac12(U+U^\dag)|n\ra$ was used.
Eq.~(\ref{mod-e-2moment-check1}) then follows from
$\sum_{n\,\rm occ}\la n|n\ra = $ 
${\rm Sp}[\Theta(E_{\rm lev}+0-\hat{H})-\Theta(-\hat{H}_0)] = B$
where the vacuum subtraction is considered explicitly and
$B=1$ denotes the baryon number \cite{Diakonov:1996sr}. 
(${\rm Sp}$ is the functional trace which can be saturated by 
respectively ${\rm Sp}[\dots]=\sum_{n\,\rm all}\la n|\dots|n\ra$ or 
$\sum_{n_0\,\rm all}\la n_0|\dots|n_0\ra$.)

For $\beta=0$ the QCD sum rule (\ref{e-2moment}) would hold ``literally''
in the $\chi$QSM. However, in the model the equations of motions are
modified compared to QCD and one cannot expect $\beta=0$ in 
(\ref{mod-e-2moment-check1},~\ref{mod-e-2moment-check1a}).
Instead, the modified equations of motion in the $\chi$QSM suggest to
interpret $\beta M$ (in the chiral limit) as the effective mass of model 
quarks bound in the soliton field.  
(One cannot expect $\beta=1$ either, which would imply an effective mass $M$.
It is jargon to refer to $M$ as mass, strictly speaking $M$ is a dimensionful
coupling of the fermion fields to the chiral background field $U$.)

%
%

\subsection{\boldmath Calculation of $(e^u+e^d)(x)$}
\label{Subsec-evaluate-e}

\paragraph{Interpolation formula.}
The approximation referred to as {\sl interpolation formula} 
\cite{Diakonov:1996sr} consists in exactly evaluating the contribution from 
the discrete level to $(e^u+e^d)(x)$ in Eq.~(\ref{mod-e-occ}), and in 
estimating the continuum contribution as follows. 
One rewrites the continuum contribution in terms of the Feynman propagator
in the static background soliton field $U$ and expands it in powers
of gradients of the $U$-field, keeping the leading term(s) only 
%
%
\footnote{
	It should be noted that this is not a strict expansion in gradients
	of the $U$-field. The dimensionless parameter characterizing this 
	expansion is $1/(MR_{sol})$. Since the soliton solution is given 
	for $MR_{sol}=1$, see Eq.~(\ref{arctan-profile}), such a strict 
	expansion is not defined \cite{Diakonov:1996sr}. In the following 
 	higher orders in the gradient expansion will be considered merely 
 	in order to study the $UV$-behaviour of the model expressions.}.
%
%
The interpolation formula yields exact results in three limiting cases: 
(i) low momenta, $|\nabla U|\ll M$, 
(ii) large momenta, $|\nabla U|\gg M$, 
(iii) any momenta but small pion field, $|\log U|\ll 1$.
One can expect that it yields useful estimates also in the general case. 
Indeed, it has been observed that estimates based on the interpolation formula
agree with results from exact (and numerically much more involved) calculations
within $10\%$ \cite{Diakonov:1996sr,Diakonov:1997vc,Weiss:1997rt}.

\paragraph{Discrete level contribution.}
The Hamiltonian $\hat{H}$ (\ref{Hamiltonian}) commutes with the parity 
operator $\hat{\pi}$ and the grand-spin operator $\hat{\bf K}$, defined 
as the sum of the total quark angular momentum and isospin operator. 
The discrete level occurs in the $K^\pi = 0^+$ sector of the 
Hamiltonian $\hat{H}$ (\ref{Hamiltonian}).  In the notation of 
Ref.~\cite{Diakonov:1996sr} the discrete level contribution reads
\be\label{discrete-level}
	(e^u+e^d)(x)_{\rm lev} = N_c\Mn\,
	\la{\rm lev}|\gamma^0\delta(x\Mn-E_{\rm lev}-\hat{p}^3)|{\rm lev}\ra 
	=  N_c\Mn\,\int\limits_{|x\Mn-E_{\rm lev}|}^\infty\!\!\!
	\frac{\di k}{2k}\biggl(h(k)^2-j(k)^2\biggr) \;, \ee
where $h(k)$ and $j(k)$ are the radial parts of respectively the upper and 
lower component of the discrete level wave function in momentum-space,
$\Phi_{\rm lev}({\bf p})=\la{\bf p}|{\rm lev}\ra$, 
see \cite{Diakonov:1996sr} for details. 

\paragraph{Dirac continuum contribution.}
The two equivalent expressions, Eqs.~(\ref{mod-e-occ},~\ref{mod-e-non}), 
allow to compute the contribution of the continuum states to $(e^u+e^d)(x)$
in two different ways
\ba
	(e^u+e^d)(x)_{\rm cont} 
	&=&     N_c \Mn \sum\limits_{E_n < 0} \la n|\gamma^0
	\delta(x\Mn-\hat{p}^3-E_n) |n\ra\label{mod-e-cont-occ}\\
	&=& -\, N_c \Mn \sum\limits_{E_n > 0} \la n|\gamma^0
	\delta(x\Mn-\hat{p}^3-E_n) |n\ra\label{mod-e-cont-non} \;.\ea
The expressions (\ref{mod-e-cont-occ}) and (\ref{mod-e-cont-non}) 
can be rewritten by means of the Feynman propagator in the static 
background pion field as (see \cite{Diakonov:1996sr})
\ba\label{gradient}
	(e^u+e^d)(x)_{\rm cont} 
	&=&
	\sum\limits_{n=0}^\infty(e^u+e^d)(x)_{\rm cont}^{(n)}\;,\nonumber\\
	(e^u+e^d)(x)_{\rm cont}^{(n)}
	&=& {\rm Im}\,N_c\Mn\,\int\!\!\frac{\di^4 p}{(2\pi)^4}\;
	\delta(p^0+p^3-x\Mn) 
	\;{\rm tr}\biggl[ \,\la {\bf p}|\,(\not\!p + MU^{-\gamma_5})\nonumber\\
	&&\times 
	\frac{1}{p_0^2+{\bf\nabla}^2-M^2+i0} \biggl(-iM 
	\gamma^i\partial_iU^{-\gamma_5}\,\frac{1}{p_0^2+{\bf\nabla}^2-M^2+i0}
	\biggr)^{\!n}\, |{\bf p}\ra \,\biggr]\Biggl|_{\rm reg} \;.\;\;\ea
The vacuum subtraction is implied in (\ref{gradient}) and means 
that the same expression but with $U\to 1$ has to be subtracted,
i.e. it is relevant only for the case $n=0$.
The subscript $_{\rm reg}$ reminds that the expression might be 
UV-divergent and has to be regularized appropriately.
Closing the $p^0$-integration contour to the upper half of the complex 
$p^0$-plane yields (\ref{mod-e-cont-occ}). Closing it to the lower 
half-plane yields (\ref{mod-e-cont-non}).

\paragraph{\boldmath Gradient expansion: Zeroth order.}
Performing the traces over $\gamma$-matrices in the zeroth order 
contribution to $(e^u+e^d)(x)_{\rm cont}$ in (\ref{gradient}) yields 
\be\label{0grad-1}
	(e^u+e^d)(x)_{\rm cont}^{(0)} = B_{sol}\;A(x)\ee
with
\ba\label{0grad-1-B}
	B_{sol}&=&
	\frac12\int\!\di^3{\bf x}\,\trF\Biggl(\frac{U+U^\dag\!}{2}-1\Biggr)
	\\
\label{0grad-1-A(x)}
	A(x) &=&  N_c\Mn\,8 M\;
	{\rm Im}\int\!\!\frac{\di^4 p}{(2\pi)^4}\;
	\frac{\delta(p^0+p^3-x\Mn)}{p^2-M^2+i0}\biggl|_{\rm reg}\;.
\ea
The coefficient $B_{sol}$ is real and contains the information on the soliton
structure.  The ``$-1$'' under the flavour-trace in (\ref{0grad-1-B}) is due 
to vacuum subtraction.
The function $A(x)$ in (\ref{0grad-1-A(x)}) is well-defined within some
appropriate regularization scheme to be figured out in the following. 
$A(x)$ is an even function of $x$, provided the regularization is consistent 
with the substitutions $p^0\to -p^0$ and $p^3\to -p^3$ in (\ref{0grad-1-A(x)}).
Keeping $x\neq0$ and integrating over $p^3$ and $p^0$ (with the 
above-mentioned prescription to close the contour) one obtains
\be\label{0grad-3}
	A(x)_{\rm non/occ} 
	= -\;\frac{\Theta(\pm x)}{|x|}\;\frac{2 N_c\, M}{\pi}\;
	\int\!\frac{\di^2{\bf p}_\perp}{(2\pi)^2}\biggl|_{\rm reg} \;,\ee
with the ``$+$'' sign referring to (\ref{mod-e-cont-non}) 
and  the ``$-$'' sign referring to (\ref{mod-e-cont-occ}).
Thus $A(x)$ is quadratically divergent, and depends on whether one 
computes it by means of (\ref{mod-e-cont-occ}) or (\ref{mod-e-cont-non}).
The non-equivalence of the two ways to compute a quantity in the model, 
(\ref{mod-e-cont-occ}) and (\ref{mod-e-cont-non}), at the level of 
{\sl unregularized} model expressions is a known phenomenon 
\cite{Diakonov:1996sr,Goeke:2000wv}.
The equivalence of (\ref{mod-e-cont-occ}) and (\ref{mod-e-cont-non}) -- and 
more generally of (\ref{matrix-elements-occ}) and (\ref{matrix-elements-non}) 
-- is a basic property of the model (see footnote~\ref{footnote-equivalence}).
Therefore it is necessary to restore the equivalence of (\ref{mod-e-cont-occ})
and (\ref{mod-e-cont-non}) in the expression (\ref{0grad-3}) by means of 
a suitably chosen regularization. A regularization -- which does this  --
is a Pauli-Villars subtraction of the type
\be\label{0grad-4}
	A(x) = A(x,M) - \frac{M}{M_1}\;A(x,M_1) \;\;, \ee
where $M_1>M$ is the Pauli-Villars mass. The Pauli-Villars subtraction is 
the privileged method to regularize divergent distribution functions in the 
$\chi$QSM. This regularization preserves all general properties of 
distribution functions (QCD sum rules, positivity, etc.) 
\cite{Diakonov:1996sr}, and where necessary it restores the equivalence of
(\ref{mod-e-cont-occ}) and (\ref{mod-e-cont-non}) in the final regularized 
model expressions \cite{Diakonov:1996sr,Goeke:2000wv}.
In the regularization (\ref{0grad-4})  -- which is sufficient at {\sl this} 
stage -- both formulae (\ref{mod-e-cont-occ}) and (\ref{mod-e-cont-non}) 
yield the same result for $A(x)$, namely
\be\label{0grad-5}
	A(x) = 0 \;\;\mbox{for}\;\; x\neq 0. \ee

Instead of studying the function $A(x)$ at the point $x=0$ it is 
more convenient to consider moments of $A(x)$ defined as 
${\cal M}_n[A]=\int\di x\, x^{n-1} A(x)$. Since $A(x)$ is even in $x$, 
one needs to consider only odd moments $2k+1$ ($k=0,1,2,\dots$)
\ba\label{0grad-6}
	{\cal M}[A]^{(2k+1)} 
	&=& \frac{8\,N_c M}{\Mn^{2k}}\,{\rm Im} 
	\sum\limits_{j=0}^k\binomial{2k}{2j}
	\int\!\!\frac{\di^4 p}{(2\pi)^4}\;
	\frac{(p^0)^{2j}(p^3)^{2k-2j}}{p_0^2-{\bf p}^2-M^2+i0}\;
	\biggl|_{\rm reg}\;. \ea
Performing a Wick rotation in Eq.~(\ref{0grad-6}), which is well defined 
for all moments under the regularization (\ref{0grad-4}), one obtains
\be\label{0grad-7}
	{\cal M}[A]^{(2k+1)} = -\;\frac{8\,N_cM}{\Mn^{2k}}
	\sum\limits_{j=0}^{k}\binomial{2k}{2j}\,(-1)^j
	\int\!\!\frac{\di^4 \pE}{(2\pi)^4}\;
	\frac{(\pE^0)^{2j}(\pE^3)^{2k-2j}}{\pE^2+M^2} \biggl|_{\rm reg}\;. \ee
Using 4-D spherical coordinates 
$\pE^\mu=q\,(\cos\psi\, , \,\sin\psi\,\sin\theta\cos\phi\, , \,\sin\psi\,\sin\theta\sin\phi\, , \,\sin\psi\,\cos\theta)$
and substituting $\kappa\equiv q^2$ one has
\ba\label{0grad-8}
&&	M[A]^{(2k+1)}	
	=
	-\;\frac{8\,N_cM\;a_k}{2(2\pi)^4\Mn^{2k}}\; \int\limits_0^\infty\!
	\frac{\di\kappa\;\kappa^{k+1}}{\kappa+M^2} \biggl|_{\rm reg}\;,	
	\nonumber\\
&&	a_k  \equiv \sum\limits_{j=0}^k\binomial{2k}{2j}\,(-1)^j
	\int\limits_0^{2\pi}\!\di\phi
	\int\limits_0^\pi\!\di\theta
	\int\limits_0^\pi\!\di\psi\;\sin\theta\;\sin^2\psi\;(\cos\psi)^{2j}\;
	(\sin\psi\,\cos\theta)^{2k-2j} = 2\pi^2\,\delta_{k0} \;.\ea
I.e. only $k=0$ contributes which means that only the first moment of $A(x)$
is non-zero
\be\label{0grad-9}
	M[A]^{(1)} =  -\;\frac{N_cM}{2\,\pi^2}\;\int\limits_0^\infty
	\!\frac{\di\kappa\;\kappa}{\kappa+M^2} \biggl|_{\rm reg} \;, \;\;\;
	M[A]^{(n)} = 0 \;\mbox{for $n\ge 2$.} \ee

The regularization prescription (\ref{0grad-4}) removes the leading quadratic
divergence in the integral in Eq.~(\ref{0grad-9}), but leaves a logarithmic 
one unregularized. However, e.g., a twofold Pauli-Villars subtraction
\be\label{0grad-10}
	A(x) = A(x,M) - \alpha_1 A(x,M_1) - \alpha_2 A(x,M_2) \ee
with 
\be\label{0grad-11}
	\alpha_1 =  \frac{M}{M_1}\;\frac{M_2^2-M^2}{M_2^2-M_1^2} \;\;, \;\;\;
	\alpha_2 = -\frac{M}{M_2}\;\frac{M_1^2-M^2}{M_2^2-M_1^2} \;\;, \;\;\; 
	M_2 > M_1 > M\;\;
\ee
is sufficient to make the first moment of $M[A]^{(1)}$ finite. In the limit 
$M_2\to\infty$ one recovers the previous regularization (\ref{0grad-4}).
It is important to note that the modification (\ref{0grad-10},~\ref{0grad-11}) 
of the previous regularization (\ref{0grad-4}) still preserves the property 
(\ref{0grad-5}).  Thus, $(e^u+e^d)(x)_{\rm cont}^{(0)}(x)$ satisfies
\be\label{0grad-12}
	(e^u+e^d)(x)_{\rm cont}^{(0)}(x) = 0\;\;\mbox{for $x\neq 0$,}\;\;\;
	\int\di x\;x^{n-1}(e^u+e^d)(x)_{\rm cont}^{(0)}(x) 
	= C\,\delta_{n1}\;, \ee
i.e. $(e^u+e^d)(x)_{\rm cont}^{(0)}(x)$ is proportional to a $\delta$-function
at $x=0$. What remains to be done is to compute the coefficient $C$ of the 
$\delta$-function in Eq.~(\ref{0grad-12}).

The two Pauli-Villars subtractions in Eq.~(\ref{0grad-10}) introduce two 
parameters, $M_1$ and $M_2$ in Eq.~(\ref{0grad-11}), which have to be fixed.
E.g. one could first fix $M_1$ by regularizing the logarithmically UV-divergent
model expression for the pion decay constant $f_\pi$ 
\be\label{0grad-13}
	f_\pi^2=4N_c\int\!\frac{\di^4\pE}{(2\pi)^4}\,\frac{M^2}{(\pE^2+M^2)^2}
	\biggl|_{\rm reg} \;,\ee
such that it gives the experimental value $f_\pi = 93\,{\rm MeV}$.
Then one could fix $M_2$, e.g., by means of the quark vacuum condensate
which is given in the effective theory (\ref{eff-theory}) by the 
quadratically divergent expression \cite{Christov:1995vm}
\be\label{0grad-14}
	\la{\rm vac}|(\bar\psi_u\psi_u+\bar\psi_d\psi_d)|{\rm vac}\ra
	= -8 N_c \int\!\frac{\di^4 \pE}{(2\pi)^4}\;\frac{M}{\pE^2+M^2}
	\biggl|_{\rm reg} \; . \ee
Two subtractions analogue to (\ref{0grad-10}) are required to regularize
(\ref{0grad-14}).
In this way the free parameters $M_1$ and $M_2$ are fixed.\footnote{
	\label{footnote-2}
	For $f_\pi^2$ in (\ref{0grad-13}) a single subtraction,
	$f_\pi^2 \equiv f_\pi^2(M)-(M/M_1)^2f_\pi^2(M_1)$ with 
	$M_1=556\,{\rm MeV}$ is required.
	For the quark condensate (\ref{0grad-14}) one needs two subtractions 
	analogue to (\ref{0grad-10},~\ref{0grad-11}) with 
	$M_2=(9.1\pm 5.7)\,{\rm GeV}$ in order to reproduce the
	phenomenological value $[(280\pm 30)\,{\rm MeV}]^3$ 
	\cite{Gasser:1982ap}.
	The first Pauli-Villars mass $M_1$ is of the order of magnitude of 
	the ``natural cutoff'' $\rho_{\rm av}^{-1}\approx 600\,{\rm MeV}$ 
	of the effective theory (\ref{eff-theory}).
	The much larger second Pauli-Villars mass $M_2$ -- in some sense 
	merely introduced as a ``technical device'' to remove the ``residual 
	(logarithmic) divergence'' left after the subtraction of the leading 
	(quadratic) divergence -- has no physical meaning.
	In non-renormalizable theories (such as the $\chi$QSM) the cutoff 
	(here $M_1\sim\rho_{\rm av}^{-1}$) has a physical meaning and shows 
	up in final expressions. In renormalizable theories the result 
	for the quadratically divergent integral (\ref{0grad-9}) would 
	be proportional to $M^2$, see e.g.\  \cite{Delbourgo:2002rh}.}
It is important to stress that the parameters $M_1$ and $M_2$ are fixed in the
vacuum- and in the meson-sector of the effective theory (\ref{eff-theory}).
In this sense the $\chi$QSM, i.e. the baryon sector of the effective theory 
(\ref{eff-theory}), produces parameter-free results.

It is interesting to observe that the coefficient $C$ can directly be
expressed in terms of the quark condensate (\ref{0grad-14}), such that
\be\label{0grad-15}
	C = B_{sol}\,\la{\rm vac}|(\bar\psi_u\psi_u+\bar\psi_d\psi_d)
	|{\rm vac}\ra\ee
with $B_{sol}$ defined in (\ref{0grad-1-B}).
In Eq.~(\ref{0grad-15}) any details on the regularization have disappeared.
However, the explicit demonstration of the existence of a regularization 
prescription, which preserves the properties in Eq.~(\ref{0grad-12}), 
was a crucial step in the derivation of (\ref{0grad-15}).

\paragraph{\boldmath Gradient expansion: Higher orders.}
Taking the trace over $\gamma$- and flavour-matrices in the expression 
for the first order term in the expansion (\ref{gradient}), 
$(e^u+e^d)(x)_{\rm cont}^{(1)}$, one obtains
\be\label{1grad}
	(e^u+e^d)(x)_{\rm cont}^{(1)} =
 	{\rm Im}\,N_c\Mn\,(-i\,8M)\int\!\!\frac{\di^4 p}{(2\pi)^4}\;
 	\frac{\delta(p^0+p^3-x\Mn)}{(p^2-M^2+i0)^2}\,p^k \la {\bf p}|\,
	\nabla^k\cos P(\hat{r},m_\pi)|{\bf p}\ra\biggl|_{\rm reg}
	 = 0\;,\ee
since $\la {\bf p}|\nabla^k\cos P(\hat{r},m_\pi)|{\bf p}\ra=$
$\int\di^3{\bf x}\,\nabla^k\cos P(r,m_\pi)=0$. A lengthy calculation 
yields for the second order contribution in (\ref{gradient}) the result
\ba\label{2grad-1}
	(e^u+e^d)(x)_{\rm cont}^{(2)} &=& {\rm Im}\,\frac{N_c\Mn M}{i 8\pi^2}
	\int\!\frac{\di\nu}{2\pi}\,e^{i\nu x\Mn}
	\int\limits_0^1\di\alpha\int\limits_0^\alpha\di\beta
	\int\!\!\di^3{\bf x}\;\trF\,{\cal A}\nonumber\\
	{\cal A} &=& U^\dag({\bf x}-\alpha\nu{\bf e^3})
	[\nabla^kU({\bf x}-\beta\nu{\bf e^3})][\nabla^kU^\dag({\bf x})]
	+ (U\leftrightarrow U^\dag) \;, \ea
which does not depend on how contours are closed, i.e. the two ways,
(\ref{mod-e-cont-occ}) and (\ref{mod-e-cont-non}), yield the same result.
In (\ref{2grad-1}) terms are neglected which contain three or more derivatives 
acting on the $U$-fields. In a strict expansion in the number $n$ of $U$-field
gradients these terms have to be considered in higher orders $n\ge 3$.
The result (\ref{2grad-1}) is real and UV-finite. Still, it has to be 
regularized according to the prescription (\ref{0grad-10},~\ref{0grad-11}).
In an exact evaluation of the continuum contribution it would be, of course, 
not possible to pick up the divergent contribution from the zeroth order
in the gradient expansion (\ref{gradient}) and regularize only that.
Since $(e^u+e^d)(x)_{\rm cont}^{(2)} \propto M$, the application of the 
regularization (\ref{0grad-10},~\ref{0grad-11}) to (\ref{2grad-1}) yields 
\be\label{2grad-2}
	(e^u+e^d)(x)_{\rm cont}^{(2)} = 0 \;.\ee
All higher orders $n\ge3$ in the gradient expansion (\ref{gradient}) are 
UV-finite. 

\paragraph{Intermediate summary.}
In \cite{Goeke:2000wv} it was shown that -- if it occurs -- the 
non-equivalence of (\ref{mod-e-cont-occ}) and (\ref{mod-e-cont-non})
at the level of unregularized model expressions only shows up in the 
lowest UV-divergent order(s) of the expansion (\ref{gradient}). 
Thus, the results of this section show that the regularization
(\ref{0grad-10},~\ref{0grad-11}) (i) consistently regularizes the
continuum contribution to $(e^u+e^d)(x)$, and (ii) ensures the 
equivalence of (\ref{mod-e-cont-occ}) and (\ref{mod-e-cont-non}).

The final regularized model expression for the continuum contribution 
consists of a $\delta$-function at $x=0$ with a coefficient proportional
to the quark condensate and the factor $B_{sol}$ in Eq.~(\ref{0grad-1-B})
which encodes the information on the nucleon (i.e.\  soliton) structure
\be\label{grad-final}
	(e^u+e^d)(x)_{\rm cont} =  C\;\delta(x)\;,\;\;\; C = B_{sol}\,
	\la{\rm vac}|(\bar\psi_u\psi_u+\bar\psi_d\psi_d)|{\rm vac}\ra\;.\ee

The existence of the $\delta$-function is a feature of the model
(with the Pauli-Villars regularization method).
The fact that the continuum contribution consists of no regular part 
but the $\delta$-function only has to be considered as a peculiarity 
of the approximation (interpolation formula) used.

%
%
\begin{figure*}
\begin{tabular}{ccc}
	\includegraphics[width=7cm,height=3.65cm]{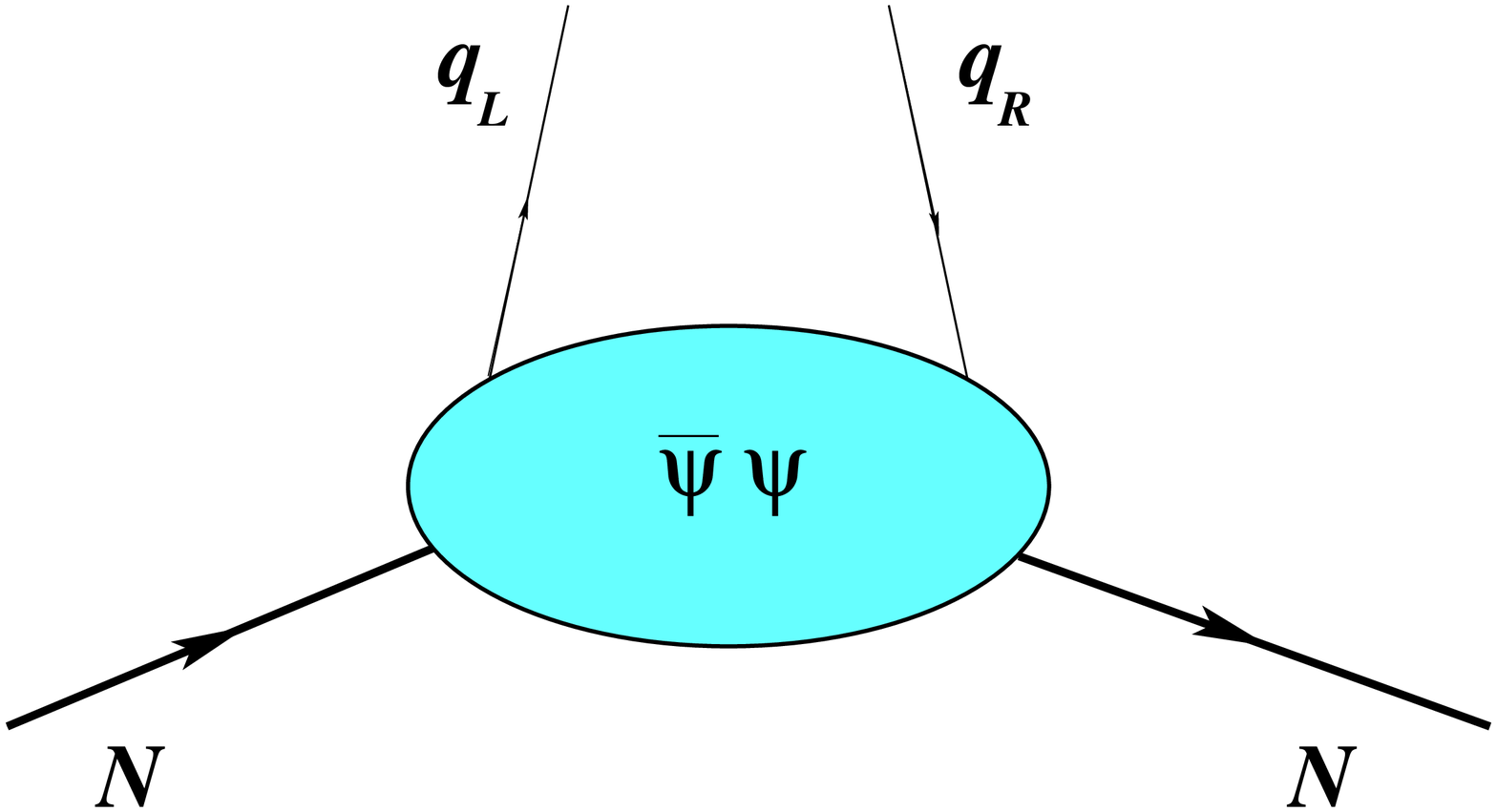} & &
	\includegraphics[width=7cm,height=3.65cm]{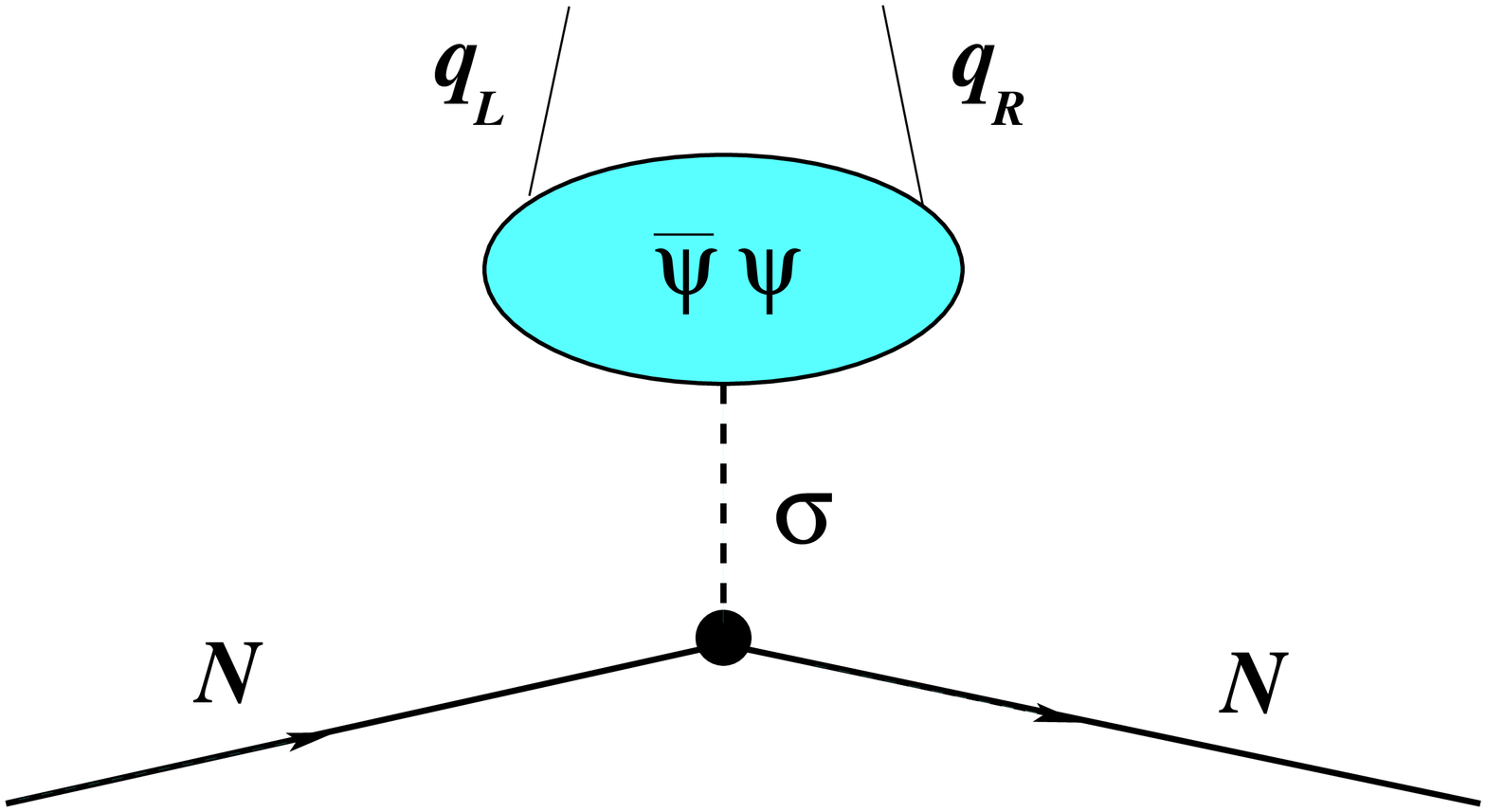}\cr
	{\bf a} & $\phantom{XXX}$& {\bf b} 
\end{tabular}
	\caption{
	{\bf a.} The oversimplified partonic interpretation of $e^q(x)$.
	{\bf b.} The symbolic diagrammatic interpretation of the $\delta(x)$ 
	contribution to $e^a(x)$ as suggested by Eq.~(\ref{grad-final}). 
	The quark-lines ``carry the fraction $x=0$ of the nucleon momentum'',
	see text.}
\end{figure*}
%
%

\paragraph{How to interpret a $\delta$-function?}
A $\delta(x)$-contribution to a distribution function has no partonic 
interpretation. The observation in (\ref{grad-final}), however, 
suggests an intuitively appealing ``interpretation''.

Oversimplifyingly $e^a(x)\di x$ can be interpreted as taking out of the 
nucleon in the infinite momentum frame, e.g., a left-handed good light-cone
quark component which carries between $x$ and $x+\di x$ of the nucleon 
momentum, and then reinserting a right-handed bad light-cone quark 
component with the same momentum back into the nucleon, see Fig.~1a.
(The good and bad quark light-cone degrees of freedom are strictly
speaking defined in the light-cone quantization. Good means independent
degrees of freedom, the bad quark degrees of freedom are composites of 
good quark and gluon degrees of freedom. Considering this one obtains the
correct partonic interpretation of $e^a(x)$ as a quark-gluon-quark
correlation function \cite{Jaffe:1991kp,Jaffe:1991ra}.)

Does it make sense to pick up hereby a quark (or antiquark) which carries 
the fraction $x=0$ of the nucleon momentum, i.e. which is at rest with 
respect to the fast moving nucleon? 
Eq.~(\ref{grad-final}) suggests that such a quark (or antiquark) is picked
up from the vacuum, which to a certain extent is present also inside the 
nucleon. 
It should be stressed that one does not deal with a disconnected diagram.
The factor $B_{sol}$ shows that the nucleon line and the vacuum blob are 
connected by the exchange of a resonance with the quantum numbers of the 
sigma meson, see the symbolic diagram in Fig.~1b.
 
It would be interesting to see whether such an ``interpretation'' 
could be confirmed by observations analogue to (\ref{grad-final}) 
in other models.

\subsection{\boldmath Discussion of the results for $(e^u+e^d)(x)$}
\label{Subsec-discuss-results}

The final result for $(e^u+e^d)(x)$ from the interpolation formula
is the contribution of the discrete level (\ref{discrete-level}), 
(already the total result for $x\neq0$) and the continuum contribution 
(\ref{grad-final}) consisting of a $\delta(x)$-function.
It should be noted that there is no freedom to also regularize the UV-finite 
discrete level contribution in the Pauli-Villars regularization method. 
This contribution must not be regularized fore otherwise the variational
problem of minimizing the soliton energy in (\ref{soliton-energy}) has no 
solution, i.e. no soliton exists \cite{Weiss:1997rt}.

Fig.~2a shows the final results for $(e^u+e^d)(x)$ and 
$(e^{\bar u}+e^{\bar d})(x)$ (no effort is made to 
indicate the $\delta$-function at $x=0$).
It is instructive to compare $(e^u+e^d)(x)$ to $(f_1^u+f_1^d)(x)$ in 
the model. Both are of the same order in the large $N_c$-limit and become 
equal in the non-relativistic limit (see Sec.~\ref{App-non-rel-limit}).
For the quarks one observes that $(f_1^u+f_1^d)(x)$ is about 2-3 times larger 
than $(e^u+e^d)(x)$, while the corresponding antiquark distributions are of 
a similar magnitude, see Figs.~2b and 2c.

%
\begin{figure*}
\begin{tabular}{ccc}
\includegraphics[width=5.7cm,height=5.7cm]{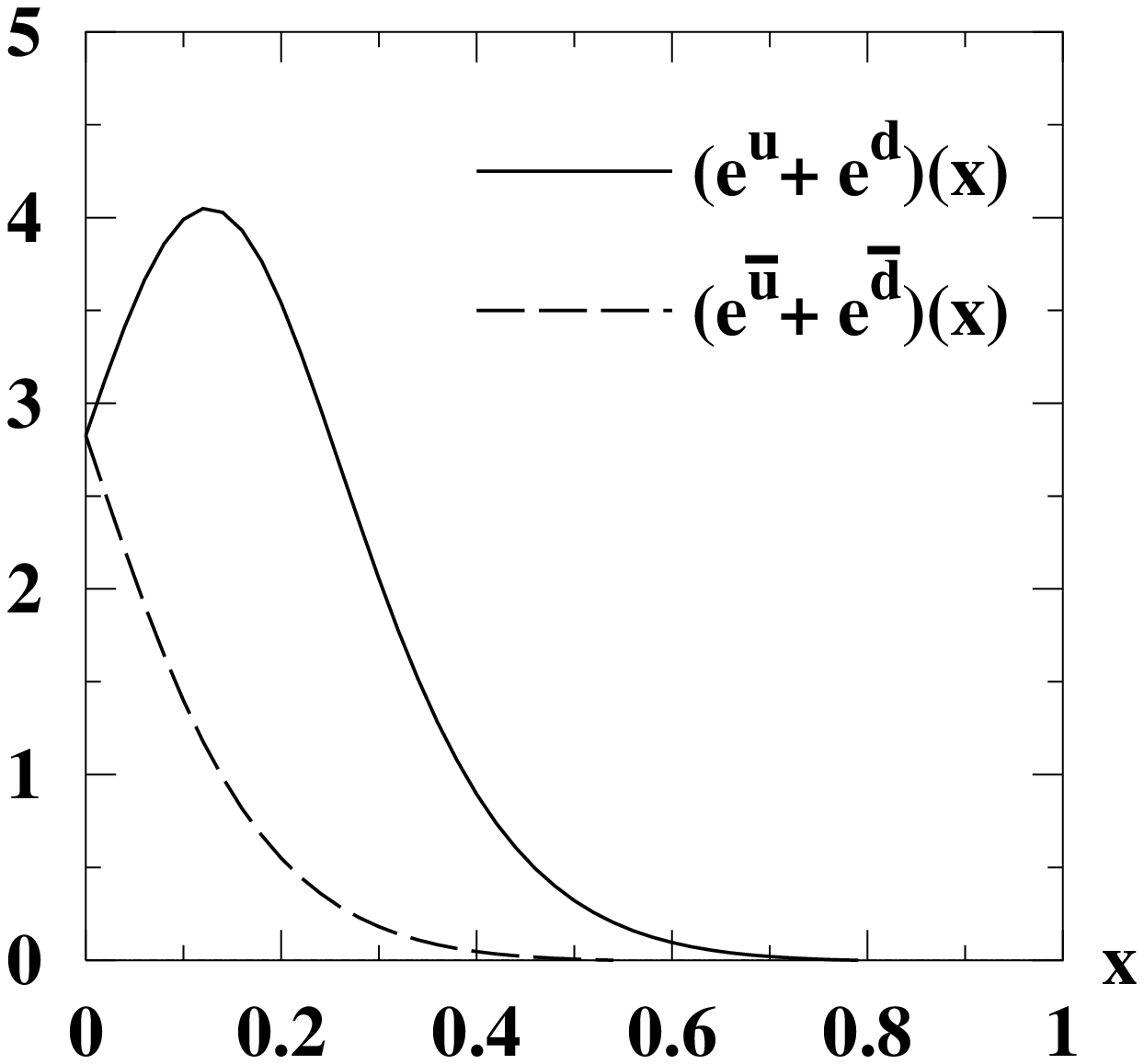}    & 
\includegraphics[width=5.7cm,height=5.7cm]{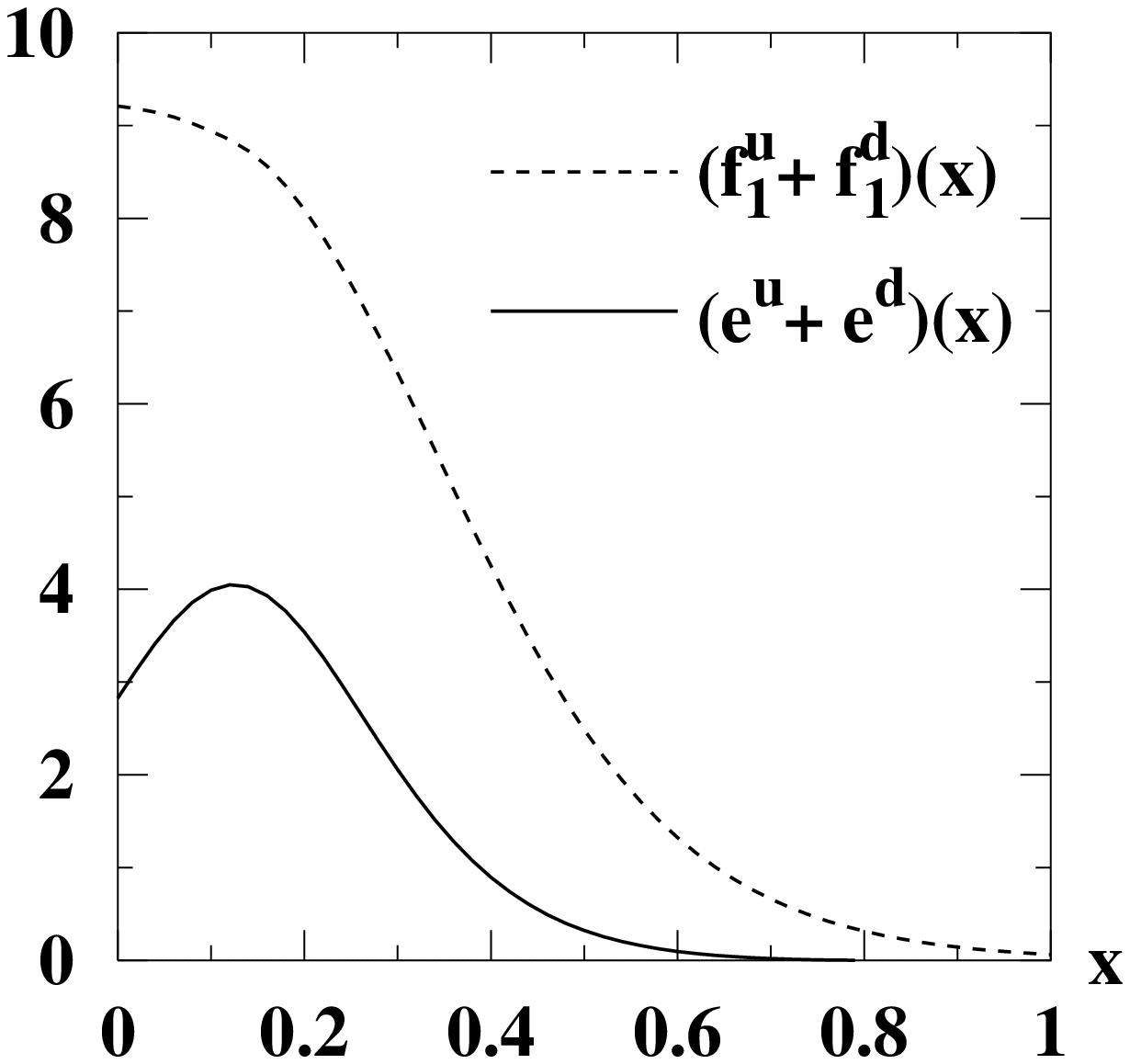} & 
\includegraphics[width=5.7cm,height=5.7cm]{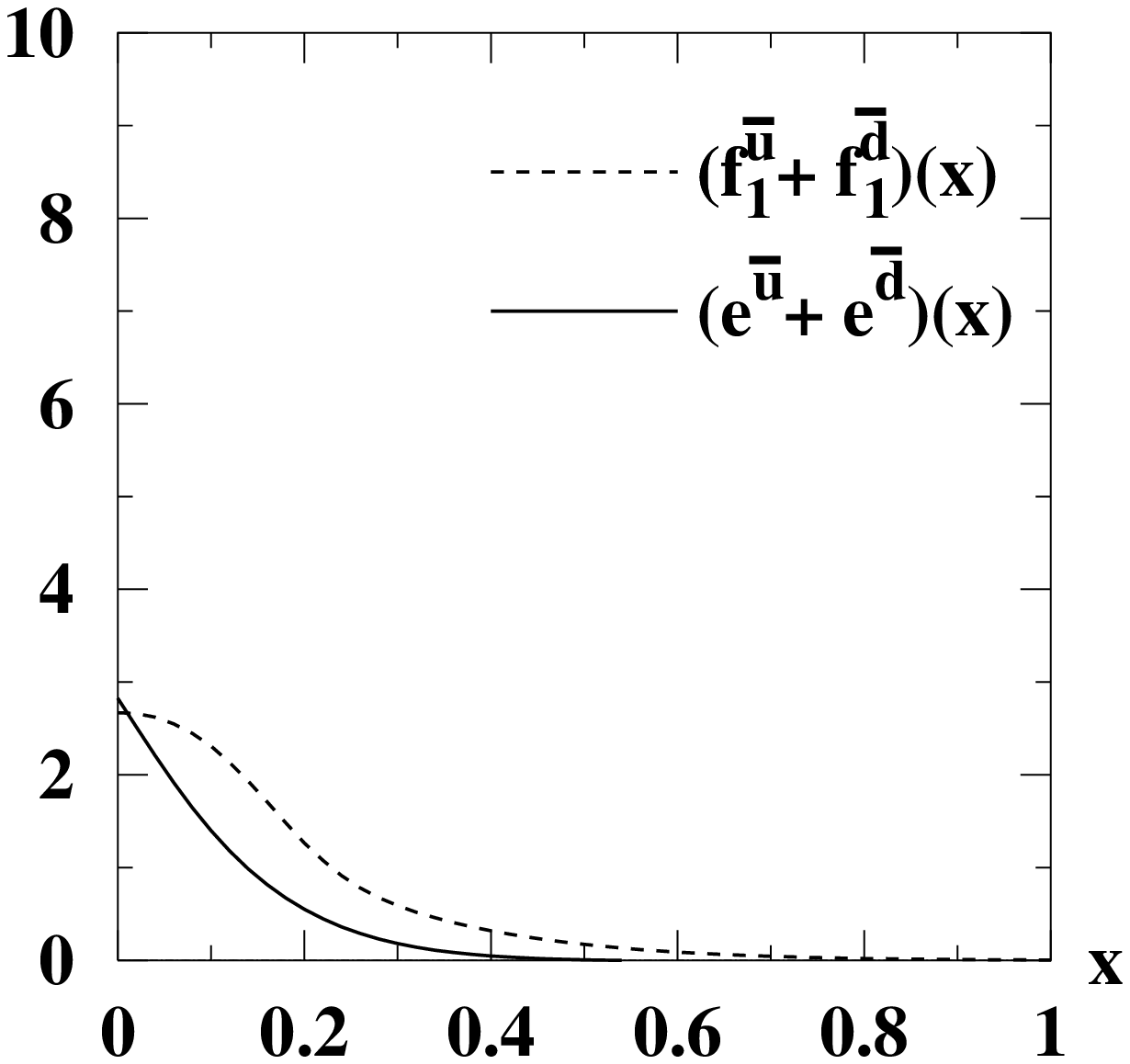} \cr
{\bf a} & 
{\bf b} & 
{\bf c} 
\end{tabular}
	\caption{
	{\bf a.}
	The regular part of the chirally odd twist-3 flavour-singlet quark 
	and antiquark distribution functions, $(e^u+e^d)(x)$ and 
	$(e^{\bar u}+e^{\bar d})(x)$, at a low normalization point 
	of the model of about $600\,{\rm MeV}$ vs.~$x$.
	At $x=0$ there is a $\delta(x)$-contribution.
	{\bf b.} 
	The chiral quark-soliton model results for $(e^u+e^d)(x)$
	(computed here) and $(f_1^u+f_1^d)(x)$ (from \cite{Weiss:1997rt})
	at a low scale of about $600\,{\rm MeV}$ vs.\ $x$. 
	{\bf c.} 
	The same as Fig.~2b but for antiquarks.}
\end{figure*}
%

In order to compute the coefficient $C$ of the $\delta(x)$-function in 
Eq.~(\ref{grad-final}), one has to evaluate $B_{sol}$ (\ref{0grad-1-B}).
For the physical situation with $m_\pi = 140\,{\rm MeV}$ one obtains
\be\label{discuss-1}
	B_{sol} =
	\frac12\int\!\di^3{\bf x}\,\trF\biggl(\frac{U+U^\dag\!}{2}-1\biggr)
	= \int\!\di^3{\bf x}\,\biggl(\cos P(r,m_\pi)-1\biggr)
	= - 17.2\;R^3_{sol} \;.\ee
In the chiral limit $m_\pi\to 0$ the integral in the expression for $B_{sol}$ 
in (\ref{0grad-1-B}) can be evaluated in an elementary way yielding
$B_{sol}=-2\sqrt{2}\,\pi^2R^3_{sol}=-27.9\,R^3_{sol}$.
I.e. the coefficient $C$ is by about $40\%$ increased in the chiral limit.
This demonstrates the importance of considering this quantity in the physical
situation with finite $m_\pi$. Numerically one obtains for 
$R_{sol}=M^{-1} = (350\,{\rm MeV})^{-1}$ and $m_\pi=140\,{\rm MeV}$ the result
\be\label{discuss-2}
	C = (9.1 \pm 2.8) \;, \ee
where the error is due to the uncertainty of the vacuum condensate 
$(-280\pm 30)^3{\rm MeV}^3$ \cite{Gasser:1982ap}.
For the first moment of $(e^u+e^d)(x)$ one thus obtains 
\be\label{discuss-3}
	\int\limits_{-1}^1\!\!\di x\;(e^u+e^d)(x) = 
	\int\limits_{-1}^1\!\!\di x\;(e^u+e^d)(x)_{\rm lev}+
	\int\limits_{-1}^1\!\!\di x\;(e^u+e^d)(x)_{\rm cont} =
	1.6 + (9.1\pm 2.8) = (10.7\pm 2.8) \ee
in good agreement with (\ref{e-1moment-b}). In order to obtain 
$\sigmaPiN$ from Eq.~(\ref{discuss-3}) it is convenient to use  
the Gell-Mann--Oakes--Renner relation 
\be\label{Gell-Mann--Oakes--Renner-relation}
	m_\pi^2 f_\pi^2 = -\,m\;\la{\rm vac}|
	(\bar\psi_u\psi_u+\bar\psi_d\psi_d)|{\rm vac}\ra \;,\ee
which holds in the effective theory (\ref{eff-theory}) \cite{Christov:1995vm}
and allows to eliminate the uncertainty from the phenomenological value of the
vacuum condensate in the continuum contribution to $\sigmaPiN$
\be\label{discuss-4}
	(\sigmaPiN)_{\rm cont} = m\int_{-1}^1\di x\;(e^u+e^d)(x)_{\rm cont} 
	= m\,C = - m_\pi^2 f_\pi^2 \;B_{\rm sol} = 67.8\,{\rm MeV}. \ee
In order to obtain the contribution of the discrete level to $\sigmaPiN$
one can use (\ref{Gell-Mann--Oakes--Renner-relation}) to obtain a consistent
value for the current quark mass $m = (8.2\pm 2.5)\,{\rm MeV}$.
This yields $(\sigmaPiN)_{\rm lev} = (13.2\pm 4.1)\;{\rm MeV}$
and the total result is $\sigmaPiN = (81\pm 4)\;{\rm MeV}$. 
There is no point in keeping track of the error in this case, 
since it is smaller than the accuracy of the interpolation formula (which 
was found to be about $\pm 10\%$ whenever it was checked quantitatively). 
Thus one obtains
\be\label{discuss-5}
	\sigmaPiN =  81\;{\rm MeV} \;. \ee
The result (\ref{discuss-4}) is about $30\%$ larger than former exact results
from the $\chi$QSM which, however, have been calculated with a different 
(proper-time) regularization \cite{Diakonov:1988mg,Kim:1995hu}.
Considering that $\sigmaPiN$ is quadratically divergent and thus rather 
strongly sensitive to regularization, the result in (\ref{discuss-5}) is in 
good agreement with the results of Refs.~\cite{Diakonov:1988mg,Kim:1995hu}. 
Worthwhile mentioning is that all model numbers -- from Eq.~(\ref{discuss-5}) 
and from Refs.~\cite{Diakonov:1988mg,Kim:1995hu} -- are consistent 
with the phenomenological value for $\sigmaPiN$ in Eq.~(\ref{sigmaPiN})
within $(10-30)\%$.

In QCD -- as mentioned in Sec.~\ref{Sec-e-in-theory} -- the first moment 
of $e^a(x)$ is due to the $\delta(x)$-function only. In the $\chi$QSM the
$\delta(x)$-function provides the dominant (more than $80\%$) but not the 
only contribution to the first Mellin moment of $(e^u+e^d)(x)$.
The second moment of $(e^u+e^d)(x)$ receives no contribution from the
continuum and is due to the discrete level contribution only 
\be\label{discuss-6}
	\int_{-1}^1\di x\;x\,(e^u+e^d)(x) = 0.20 \;. \ee
This result would imply that the quarks bound in the soliton field have an 
effective mass $\beta M \sim 78\,{\rm MeV}$ (in the chiral limit), see the 
discussion below (\ref{mod-e-2moment-check1},~\ref{mod-e-2moment-check1a}).

A comment is in order on an exact numerical evaluation of $(e^u+e^d)(x)$.
The distribution functions computed in the $\chi$QSM so far
were all either UV-finite or at most logarithmically divergent.
In the latter case always a single Pauli-Villars subtraction was sufficient.
The regularization prescription (\ref{0grad-10},~\ref{0grad-11}) 
precisely states how $(e^u+e^d)(x)$ can practically be regularized.
However, an exact numerical evaluation meets the problem to evaluate the 
model expressions for a Pauli-Villars mass $M_2\approx$ several ${\rm GeV}$
(see footnote~\ref{footnote-2})
In the numerical calculation the spectrum of the Hamiltonian 
(\ref{Hamiltonian}) is discretized and made finite 
(see e.g.\  \cite{Diakonov:1997vc} and references therein).
For the latter step one considers only quark momenta below some large 
numerical cutoff $\Lambda_{\rm num}$ chosen much larger than any 
other (physical or numerical) scale involved in the problem.
So far $\Lambda_{\rm num} \approx (8-10)\,{\rm GeV}$ was sufficient,
but this is the order of magnitude of the second Pauli-Villars mass $M_2$.
To compute $(e^u+e^d)(x)$ one has to choose $\Lambda_{\rm num}$ much larger
than $M_2$, which would result in an uneconomically large increase of 
computing time.

Interestingly, the singular $\delta(x)$-contribution would conceptually 
cause no problem for the numerical method of Ref.~\cite{Diakonov:1997vc}.
The descretized spectrum of the Hamiltonian (\ref{Hamiltonian}) 
yields discontinuous (distribution) functions of $x$. 
In \cite{Diakonov:1997vc} it was proposed to smear the distribution 
functions $q(x)$, i.e. to convolute them with a narrow Gaussian with an
appropriately chosen width $\gamma$ as $q(x)_{\rm smear} =$
$\frac{1}{\gamma\pi^{1/2}}\int\di x'\exp(-\frac{(x-x')^2}{\gamma^2})q(x')$.
(The smearing can be removed by a deconvolution procedure.)
This trick would turn the $\delta(x)$-contribution in $(e^u+e^d)(x)$ into 
a narrow-Gausssian with the well-defined width $\gamma$. In this way the
coefficient of the $\delta(x)$-contribution could be well determined from 
the numerical result.

Finally, a comment is in order on (\ref{sigma-Mn-relation}) which relates the
logarithmically divergent nucleon mass $\Mn$ and the quadratically divergent
$\sigmaPiN$. $\Mn$ requires a single Pauli-Villars subtraction, while 
$\sigmaPiN$ requires two subtractions.
Thus the quantities on the left-hand and right-hand side in 
(\ref{sigma-Mn-relation}) are regularized differently. In the 
Pauli-Villars regularization scheme the relation (\ref{sigma-Mn-relation})
has to be considered as formally correct modulo regularization effects. 
   (Some other ambiguities in the Pauli-Villars regularization 
   scheme were mentioned in \cite{Kubota:1999hx}.) 
In other regularization methods -- such as the proper 
time regularization -- there are no such ambiguities.

\subsection{Comparison to the bag model}

%
	\begin{figure}
	\includegraphics[width=5.7cm,height=5.7cm]{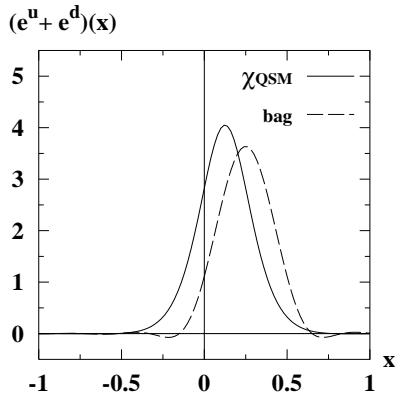}
   	\caption{
	The twist-3 distribution function $(e^u+e^d)(x)$ from the $\chi$QSM
	(solid line, obtained here) and from the bag model (dashed line,
	from Ref.~\cite{Jaffe:1991ra}) vs.~$x$. The meaning of the curves at 
	negative $x$ is explained in Eq.~(\ref{def-e}).
	The $\chi$QSM result refers to a scale of about $0.6\,{\rm GeV}$.
	The bag model result refers to a scale of about $0.4\,{\rm GeV}$.}
\end{figure}
%
%
Studies of $e^a(x)$ were also performed in the framework of the bag model 
in Refs.~\cite{Jaffe:1991ra,Signal:1997ct}. In Fig.~3 the $\chi$QSM result
for the regular part of $(e^u+e^d)(x)$ is compared to the result from the 
MIT bag model from Ref.~\cite{Jaffe:1991ra}. 
(For that the flavour-independent results for ``$e(x)$'' from 
Refs.~\cite{Jaffe:1991ra} are multiplied by the factor $N_c=3$ 
in order to compare to $(e^u+e^d)(x)$ obtained here.)
The comparison of the position of the maxima of the curves for quark 
distributions from the two models indicates that the bag model results
refer to a somehow lower scale than $0.6\,{\rm GeV}$, the scale of the
$\chi$QSM. (In \cite{Signal:1997ct} the value of $0.4\,{\rm GeV}$ was
quoted.) 
Taking this into account one concludes that both models give 
qualitatively similar results for quark distributions.

Concerning antiquarks the difference is more pronounced. However, 
the bag model description of antiquark distributions cannot be considered
as reliable. A drawback of the bag model in this context is that it yields 
negative $f_1^{\bar q}(x)$ in contradiction to the positivity requirement.

\section{\boldmath $e^a(x)$ in the non-relativistic limit}
\label{App-non-rel-limit}

In the limit of the soliton size $R_{sol}\to 0$ the expressions of the 
$\chi$QSM go into the results of the non-relativistic (``naive'') quark 
model formulated for an arbitrary number of colours $N_c$ \cite{Karl:cz}.
In this sense the limit $R_{sol}\to 0$ corresponds to the non-relativistic 
limit in the $\chi$QSM. 
This was studied in detail in Ref.~\cite{Praszalowicz:1995vi}.

As $R_{sol}\to 0$ the soliton profile (\ref{arctan-profile}) goes to zero,
and the $U$-field approaches unity. Correspondingly, the spectrum of the
Hamiltonian (\ref{Hamiltonian}) becomes more and more similar to that of 
the free Hamiltonian (\ref{free-Hamiltonian}).
Considering vacuum subtraction it is clear that the contribution of the 
continuum vanishes in this limit \cite{Praszalowicz:1995vi}, and all that 
remains is the contribution of the discrete level $|{\rm lev}\ra$. 
More precisely, as $R_{sol}\to 0$ the energy of the discrete level
$E_{\rm lev}\to M$  (so the nucleon mass, (\ref{soliton-energy}), formally 
$\Mn\to N_c E_{\rm lev}\to N_cM$), and the lower component of the Dirac-spinor
of the discrete level wave-function goes to zero \cite{Praszalowicz:1995vi}.

The limit $R_{sol}\to 0$ means that $U\to 1$ i.e.\ ${\rm log}\,U \ll 1$
which is the case (iii) in which the interpolation formula yields exact
results (see Section~\ref{Subsec-evaluate-e}).
The first feature -- in this case the vanishing of the continuum contribution
in (\ref{mod-e-cont-occ},~\ref{mod-e-cont-non}) -- can be observed in the final
result $(e^u+e^d)(x)_{\rm cont} = C\,\delta(x)$ in Eq.~(\ref{grad-final}). 
The factor $B_{sol}$ in the coefficient $C$ vanishes with $R_{sol}\to 0$ as
can be seen from its definition (\ref{0grad-1-B}) or Eq.~(\ref{discuss-1}).
Thus, in the non-relativistic limit the contribution of the $\delta$-function 
at $x=0$ vanishes because the coefficient $C$ goes to zero. 

To study the non-relativistic limit in the discrete level contribution it 
is convenient to use the first expression in Eq.~(\ref{discrete-level}).
Since only the upper component of the Dirac-spinor of the discrete level 
wave-function survives the limit \cite{Praszalowicz:1995vi}, one can replace 
$\gamma^0$ by the unity matrix, i.e.
\be\label{non-rel-lim-1}
	\limNR(e^u+e^d)(x) \longrightarrow N_c\Mn
	\la{\rm lev}|\delta(x\Mn-\hat{p}^3-E_{\rm lev})|{\rm lev}\ra \;.\ee
Next consider that $E_{\rm lev}\to M$ and $\Mn \to N_cM$ while the momenta 
of the non-relativistic quarks $|p^i|\ll M$ such that the corresponding 
operator in the $\delta$-function in (\ref{non-rel-lim-1}) can be neglected.
Using the normalization $\la{\rm lev}|{\rm lev}\ra=1$ one obtains
\be\label{non-rel-lim-2}
	\limNR(e^u+e^d)(x) \longrightarrow N_c\, 
	\delta\biggl(x-\frac{1}{N_c}\biggr) \;.\ee
The first two moments of (\ref{non-rel-lim-2}) read
\ba\label{non-rel-lim-3}
	\limNR\int\limits_{-1}^1\!\!\di x\;   (e^u+e^d)(x) = N_c  &,& 
	\limNR\int\limits_{-1}^1\!\!\di x\;x\,(e^u+e^d)(x) = 1 \;. \ea
To see that the relations (\ref{non-rel-lim-3}) are the correct
non-relativistic results for the QCD sum rules (\ref{e-2moment}) and
(\ref{e-1moment-a}) one has to consider that in the non-relativistic limit
the current quark mass $m\to M = \Mn/N_c$ and $\sigmaPiN\to \Mn$.
(The latter relation follows formally, e.g., from the Feynman-Hellmann 
theorem (\ref{sigma-Mn-relation}) with $m = M = \Mn/N_c$.)

The twist-2 unpolarized flavour-singlet distribution function 
$(f_1^u+f_1^d)(x)$ is given in the $\chi$QSM by \cite{Diakonov:1996sr}
\ba\label{non-rel-lim-4}
	(f_1^u+f_1^d)(x)  =  N_c \Mn \sum\limits_{n\,\rm occ} 
		\la n|(1+\gamma^0\gamma^3)\,\delta(x\Mn-p^3-E_n) |n\ra \;,\ea
i.e.\  the  only difference to $(e^u+e^d)(x)$ is the different 
Dirac-structure $(1+\gamma^0\gamma^3)$ instead of $\gamma^0$. 
This difference becomes irrelevant in the non-relativistic limit (where only 
the upper component of the Dirac-spinor of the discrete level wave-function 
survives) such that in this limit $(e^u+e^d)(x)$ and  $(f_1^u+f_1^d)(x)$
become equal. This argument holds also for separat flavours and allows to 
generalize
\be\label{non-rel-lim-5}
	\limNR e^q(x) = \limNR f_1^q(x) = N_q\,
	\delta\biggl(x-\frac{1}{N_c}\biggr) \;,\ee
where $N_q$ denotes the number of the respective valence quarks
(i.e. for the proton $N_u=2$ and $N_d=1$ for $N_c=3$ colours).
The result (\ref{non-rel-lim-5}) means that in the non-relativistic limit
$e^q(x)$ is given by the mass term contribution in the decomposition 
(\ref{e-decomposition}) because
\be\label{non-rel-lim-6}
	\limNR\biggl\{\frac{m_q}{\Mn}\;\frac{f_1^q(x)}{x}\biggr\}
	= \frac{M}{\Mn}  \limNR\biggl\{\frac{f_1^q(x)}{x}\biggr\}
	= \limNR f_1^q(x) \;. \ee
In the intermediate step in (\ref{non-rel-lim-6}) the 
$\delta(x-1/N_c)$-function was used to replace $x$ in the 
denominator by $1/N_c$. 
One may worry that the large value $\sigmaPiN=\Mn$ would mean a large
strangeness content of the nucleon. However, as discussed in 
\cite{Efremov:2002qh} the value $\sigmaPiN=\Mn$ correctly implies a vanishing 
strangeness contribution to the nucleon mass in the non-relativistic limit.

Thus, though phenomenologically it is not satisfactory, the non-relativistic 
picture of the twist-3 distribution function $e^q(x)$ is consistent.
In this limit the singular and pure twist-3 contributions in the decomposition
(\ref{e-decomposition}) vanish, and $e^q(x)$ is given by the mass-term. 
Thus the chirally odd nature of $e^q(x)$ arises from a ``mass insertion'' into
a quark line. Moreover, $e^q(x)$ and $f_1^q(x)$ become equal in this limit,
and are trivial $\delta$-functions concentrated at $x=1/N_c$
which means that the nucleon momentum is distributed equally among
the $N_c$ massive and non-interacting constituent quarks.

The usefulness of results of the kind (\ref{non-rel-lim-5}) is best 
illustrated by the popularity of the non-relativistic relation between
the twist-2 helicity $g_1^q(x)$ and transversity $h_1^q(x)$ distribution 
functions (here for $N_c=3$)
\be\label{non-rel-lim-7}
	\limNR h_1^q(x) = \limNR g_1^q(x) = P_q\,
	\delta\biggl(x-\frac{1}{3}\biggr) \;,\;\;\;
	P_u = \frac{4}{3} \;, \;\;\;
 	P_d = -\,\frac{1}{3} \;, \ee
which yields for the axial charges $g_A^{(3)}=\frac{5}{3}$ and $g_A^{(0)}=1$.
Though also these numbers are phenomenologically not fully satisfactory
the assumption that $h_1^a(x)=g_1^a(x)$ at some low scale is a popular guess 
to estimate effects of transversity distribution, see \cite{Barone:2001sp} 
for a review.

\section{Summary and conclusions}
\label{conclusions}

A study of the flavour-singlet twist-3 distribution function $(e^u+e^d)(x)$
in the $\chi$QSM was presented. 
It was shown that the model expressions are quadratically and logarithmically 
UV-divergent and can be regularized by the Pauli-Villars method.
The model expressions for the quark and antiquark distribution functions 
$(e^u+e^d)(x)$ and $(e^{\bar u}+e^{\bar d})(x)$ were evaluated using an 
approximation -- the interpolation formula which in general well approximates
exact model calculations.

The remarkable result is that the $\chi$QSM-expression for $(e^u+e^d)(x)$ 
contains a $\delta$-function-type singularity at $x=0$ as expected from QCD 
\cite{Efremov:2002qh}.
This result is obtained here from a non-perturbative model calculation.
Previously a $\delta(x)$-contribution in $e^a(x)$ was observed 
in a perturbative calculation in Ref.~\cite{Burkardt:2001iy}.

In the $\chi$QSM the coefficient of the $\delta$-function is proportional to 
the quark vacuum condensate. This is natural from the point of view that in 
QCD the singular contribution to $e^q(x)$ and the quark vacuum condensate 
are both the expectation values of the same local scalar operator, 
$\bar\psi_q(0)\psi_q(0)$, taken respectively in the nucleon and vacuum states.
This observation allows to make a heuristic but physically appealing 
interpretation of the $\delta(x)$-contribution.

At $x\neq 0$ the $\chi$QSM yields results for $(e^u+e^d)(x)$ similar 
to those obtained in the bag model at a comparably low scale 
\cite{Jaffe:1991ra,Signal:1997ct}.
Both models suggest that $e^a(x)$ is sizable at low scales.
The discription of $(e^u+e^d)(x)$ in the $\chi$QSM is consistent in the 
sense that the sum rules for the first and the second moment are satisfied. 
However, in the $\chi$QSM the $\delta(x)$-contribution provides the dominant 
but not the only contribution to the first moment of $(e^u+e^d)(x)$ unlike in 
QCD, and in the case of the second moment it is necessary to interpret the 
result correspondingly by introducing the notion of an effective quark mass.
It would be interesting to see whether the failure of the bag model to 
satisfy the sum rule for the second moment reported in  \cite{Jaffe:1991ra} 
could also be reinterpreted in a similar spirit.

In effective models, such as $\chi$QSM or bag model, equations of motions are 
altered compared to QCD and there is no gauge principle which would allow to 
cleanly decompose $e^a(x)$ into a $\delta(x)$-contribution, a pure twist-3 
part and a mass-term.
Therefore one cannot expect that the QCD sum rules which are derived by 
means of the QCD equations of motion are literally satisfied in such models.
Still within the models the results are consistent.

The (large-$N_c$) non-relativistic limit of $e^q(x)$ was studied on the 
basis of the $\chi$QSM expressions.
It was found that in this limit $e^q(x)=f_1^q(x)$.
The non-relativistic description of $e^q(x)$ was shown to be consistent.

The results and interpretations presented here should be reexamined in 
the more general framework of the instanton model of the QCD vacuum,
on which the $\chi$QSM is founded. This was out of the scope of 
the study presented here and will be reported elsewhere.

\begin{acknowledgments}
I would like to thank A.~V.~Efremov, K.~Goeke, P.~V.~Pobylitsa, 
M.~V.~Polyakov and C.~Weiss for many fruitful discussions.
This work has partly been performed under the contract  
HPRN-CT-2000-00130 of the European Commission.
\end{acknowledgments}

%
%
\vspace{0.5cm}
{\sl Note added in Proof:}
After this work has been completed the work \cite{Wakamatsu:2003uu} appeared, 
where the authors conclude the existence of a $\delta(x)$ contribution in 
$(e^u+e^d)(x)$ in the chiral quark-soliton model in an independent and 
complementary way.
%
%



\begin{thebibliography}{99}

\bibitem{Jaffe:1991kp}
   R.~L.~Jaffe and X.~D.~Ji, Phys.\ Rev.\ Lett.\  {\bf 67}, 552 (1991).
\bibitem{Jaffe:1991ra}
   R.~L.~Jaffe and X.~D.~Ji, Nucl.\ Phys.\ B {\bf 375}, 527 (1992).
\bibitem{Balitsky:1996uh}
   I.~I.~Balitsky, V.~M.~Braun, Y.~Koike and K.~Tanaka,
   Phys.\ Rev.\ Lett.\  {\bf 77}, 3078 (1996). 
\bibitem{Belitsky:1997zw}
   A.~V.~Belitsky and D.~Muller,
   Nucl.\ Phys.\ B {\bf 503}, 279 (1997). 
\bibitem{Koike:1997bs}
   Y.~Koike and N.~Nishiyama,
   Phys.\ Rev.\ D {\bf 55}, 3068 (1997). 
\bibitem{Belitsky:1997ay}
   A.~V.~Belitsky,
   in Proceedings of the ``31st PNPI Winter School on Nuclear and Particle 
   Physics'', St. Petersburg, Russia, 24 Feb - 2 Mar 1997,
   ed.~ V.~A.~Gordeev, pp.369-455 [arXiv:hep-ph/9703432].
\bibitem{Kodaira:1998jn}
   J.~Kodaira and K.~Tanaka,
   Prog.\ Theor.\ Phys.\  {\bf 101}, 191 (1999). 
\bibitem{Efremov:2002qh}
   A.~V.~Efremov and P.~Schweitzer, arXiv:hep-ph/0212044.
\bibitem{Burkardt:2001iy}
   M.~Burkardt and Y.~Koike,
   Nucl.\ Phys.\ B {\bf 632}, 311 (2002). 
\bibitem{Koch:pu}
   R.~Koch, Z.\ Phys.\ C {\bf 15}, 161 (1982).
\bibitem{Pavan:2001wz}
   M.~M.~Pavan, I.~I.~Strakovsky, R.~L.~Workman and R.~A.~Arndt,
   $\pi$N Newsletter {\bf 16}, 110 (2002) [arXiv:hep-ph/0111066].
\bibitem{collins}
   J.~C.~Collins,
   Nucl.\ Phys.\ B {\bf 396}, 161 (1993). \\ 
   X.~Artru and J.~C.~Collins,
   Z.\ Phys.\ C {\bf 69}, 277 (1996). 
\bibitem{muldt}
   D.~Boer and P.~J.~Mulders, Phys. Rev. \textbf{D57}, 5780 (1998). \\
   D.~Boer, R.~Jakob and P.~J.~Mulders, Phys. Lett. \textbf{B424}, 143 (1998).
   \\
   D.~Boer and R.~Tanger\-man, Phys. Lett. \textbf{B381}, 305 (1996).

\bibitem{Efremov:1998vd}
   A.~V.~Efremov, O.~G.~Smirnova and L.~G.~Tkachev,
   Nucl.\ Phys.\ Proc.\ Suppl.\  {\bf 74}, 49 (1999); 
   Nucl.\ Phys.\ Proc.\ Suppl.\  {\bf 79}, 554 (1999).
   A.~V.~Efremov {\it et al.}, 
   Czech.\ J.\ Phys.\  {\bf 49}, S75 (1999) [arXiv:hep-ph/9901216].

\bibitem{Mulders:1996dh}
   P.~J.~Mulders and R.~D.~Tangerman,
   Nucl.\ Phys.\ {\bf B461}, 197 (1996)
   [Erratum-ibid.\ {\bf B484}, 538 (1996)]. 
\bibitem{hermes}
   A.~Airapetian {\it et al.}  [HERMES Collaboration],
   Phys.\ Rev.\ Lett.\  {\bf 84}, 4047 (2000); 
   H.~Avakian  [HERMES Collaboration],
   Nucl.\ Phys.\ Proc.\ Suppl.\  {\bf 79}, 523 (1999); 
   A.~Airapetian {\it et al.}  [HERMES Collaboration],
   Phys.\ Rev.\ D {\bf 64}, 097101 (2001). 
\bibitem{Avakian-talk}
   H.~Avakian [CLAS Collaboration], talk at 9th International Conference
   on the Structure of Baryons (Baryons 2002), Newport News, Virginia, 
   3-8 March 2002.
\bibitem{Avakian:2003pk}
   H.~Avakian {\it et al.}  [CLAS Collaboration],
   arXiv:hep-ex/0301005.
\bibitem{Efremov:2001cz}
   A.~V.~Efremov, K.~Goeke and P.~Schweitzer,
   Phys.\ Lett.\ B {\bf 522}, 37 (2001)
   [Erratum-ibid.\ B {\bf 544}, 389 (2002)]. 
\bibitem{Efremov:2002sd}
   A.~V.~Efremov, K.~Goeke and P.~Schweitzer,
   Acta Phys.\ Polon.\ B {\bf 33}, 3755 (2002); 
   and
   arXiv:hep-ph/0208124.
\bibitem{Signal:1997ct}
   A.~I.~Signal, Nucl.\ Phys.\ B {\bf 497}, 415 (1997). 
\bibitem{Diakonov:1985eg}
   D.~I.~Diakonov and V.~Y.~Petrov, Nucl.\ Phys.\ B {\bf 272}, 457 (1986).
\bibitem{Diakonov:1983hh}
   D.~I.~Diakonov and V.~Y.~Petrov, Nucl.\ Phys.\ B {\bf 245}, 259 (1984).
\bibitem{Diakonov:1995qy}
   D.~I.~Diakonov, M.~V.~Polyakov and C.~Weiss,
   Nucl.\ Phys.\ B {\bf 461}, 539 (1996). 
\bibitem{Diakonov:1996sr}
   D.~I.~Diakonov, V.~Petrov, P.~Pobylitsa, M.~V.~Polyakov and C.~Weiss,
   Nucl.\ Phys.\ B {\bf 480}, 341 (1996). 
\bibitem{Diakonov:1997vc}
   D.~I.~Diakonov, V.~Y.~Petrov, P.~V.~Pobylitsa, M.~V.~Polyakov and  C.~Weiss,
   Phys.\ Rev.\ D {\bf 56}, 4069 (1997). 
\bibitem{Weiss:1997rt}
   C.~Weiss and K.~Goeke, arXiv:hep-ph/9712447.
\bibitem{Pobylitsa:1996rs}
   P.~V.~Pobylitsa and M.~V.~Polyakov,
   Phys.\ Lett.\ B {\bf 389}, 350 (1996);\\ 
   P.~V.~Pobylitsa {\it et al.},
   Phys.\ Rev.\ D {\bf 59}, 034024 (1999);\\ 
   M.~Wakamatsu and T.~Kubota,
   Phys.\ Rev.\ D {\bf 60}, 034020 (1999);\\ 
   P.~Schweitzer {\it et al.}, 
   Phys.\ Rev.\ D {\bf 64}, 034013 (2001). 
\bibitem{Goeke:2000wv}
   K.~Goeke, P.~V.~Pobylitsa, M.~V.~Polyakov, P.~Schweitzer and D.~Urbano,
   Acta Phys.\ Polon.\ B {\bf 32}, 1201 (2001). 
\bibitem{GRV+GRSV}
   M.~Gl\"uck, E.~Reya and A.~Vogt, Z.\ Phys.\ C {\bf 67}, 433 (1995).
   M.~Gl\"uck, E.~Reya, M.~Stratmann and W.~Vogelsang,
   Phys.\ Rev.\ D {\bf 53}, 4775 (1996). 
\bibitem{Balla:1997hf}
   J.~Balla, M.~V.~Polyakov and C.~Weiss,
   Nucl.\ Phys.\ B {\bf 510}, 327 (1998). 
\bibitem{Dressler:2000hc}
   B.~Dressler and M.~V.~Polyakov,
   Phys.\ Rev.\ {\bf D61}, 097501 (2000). 
\bibitem{Wakamatsu:2000ex}
   M.~Wakamatsu,
   Phys.\ Lett.\ B {\bf 487}, 118 (2000). 
\bibitem{Wakamatsu:2001fd}
   M.~Wakamatsu,
   Phys.\ Lett.\ B {\bf 509}, 59 (2001). 
\bibitem{Gasser:1990ce}
   J.~Gasser, H.~Leutwyler and M.~E.~Sainio, 
   Phys.\ Lett.\ B {\bf 253}, 252 and 260 (1991).
\bibitem{Becher:1999he}
   T.~Becher and H.~Leutwyler,
   Eur.\ Phys.\ J.\ C {\bf 9}, 643 (1999). 
\bibitem{Diakonov:1987ty}
   D.~I.~Diakonov, V.~Y.~Petrov and P.~V.~Pobylitsa,
   Nucl.\ Phys.\ B {\bf 306}, 809 (1988).\\
   D.~I.~Diakonov and V.~Y.~Petrov, JETP Lett.\  {\bf 43}, 75 (1986)
   [Pisma Zh.\ Eksp.\ Teor.\ Fiz.\  {\bf 43}, 57 (1986)].
\bibitem{Diakonov:tw}
   D.~I.~Diakonov and M.~I.~Eides, JETP Lett.\  {\bf 38}, 433 (1983)
   [Pisma Zh.\ Eksp.\ Teor.\ Fiz.\  {\bf 38}, 358 (1983)];
   A.~Dhar, R.~Shankar and S.~R.~Wadia,	Phys.\ Rev.\ D {\bf 31}, 3256 (1985). 
\bibitem{Christov:1995vm} 
   C.~V.~Christov {\it et al.},
   Prog.\ Part.\ Nucl.\ Phys.\  {\bf 37}, 91 (1996). 
\bibitem{Petrov:1998kf}
   V.~Y.~Petrov, P.~V.~Pobylitsa, M.~V.~Polyakov, 
   I.~B\"ornig, K.~Goeke and C.~Weiss, 
   Phys.\ Rev.\ D {\bf 57}, 4325 (1998);\\ 
   M.~Penttinen, M.~V.~Polyakov and K.~Goeke,
   Phys.\ Rev.\ D {\bf 62}, 014024 (2000); \\ 
   P.~Schweitzer, S.~Boffi and M.~Radici,
   Phys.\ Rev.\ D {\bf 66}, 114004 (2002). 
\bibitem{Witten:1979kh}
   E.~Witten,
   Nucl.\ Phys.\ B {\bf 160}, 57 (1979);
   E.~Witten, Nucl.\ Phys.\ B {\bf 223}, 433 (1983).
\bibitem{Diakonov:1988mg}
   D.~I.~Diakonov, V.~Y.~Petrov and M.~Prasza\l owicz,
   Nucl.\ Phys.\ B {\bf 323}, 53 (1989).
\bibitem{Kim:1995hu}
   H.~C.~Kim, A.~Blotz, C.~Schneider and K.~Goeke,
   Nucl.\ Phys.\ A {\bf 596}, 415 (1996). 
\bibitem{Gasser:1982ap}
   J.~Gasser and H.~Leutwyler, Phys.\ Rept.\  {\bf 87}, 77 (1982).
\bibitem{Delbourgo:2002rh}
   R.~Delbourgo and M.~D.~Scadron,
   Mod.\ Phys.\ Lett.\ A {\bf 17}, 209 (2002). 
\bibitem{Kubota:1999hx}
   T.~Kubota, M.~Wakamatsu and T.~Watabe,
   Phys.\ Rev.\ D {\bf 60}, 014016 (1999). 
\bibitem{Karl:cz}
   G.~Karl and J.~E.~Paton,
   Phys.\ Rev.\ D {\bf 30}, 238 (1984).
\bibitem{Praszalowicz:1995vi}
   M.~Prasza\l owicz, A.~Blotz and K.~Goeke,
   Phys.\ Lett.\ B {\bf 354}, 415 (1995). 
\bibitem{Barone:2001sp}
   V.~Barone, A.~Drago and P.~G.~Ratcliffe,
   Phys.\ Rept.\  {\bf 359}, 1 (2002). 
\bibitem{Wakamatsu:2003uu}
   M.~Wakamatsu and Y.~Ohnishi,
   {\sl preprint} OU-HEP-433, arXiv:hep-ph/0303007 (2003).

\end{thebibliography}
\end{document}